\documentclass[a4paper,runningheads]{llncs}

\newif\ifblind
\blindfalse

\newif\ifdraft
\drafttrue

\newif\iflong
\longfalse

\usepackage[T1]{fontenc}
\usepackage{times}
\usepackage[scaled=0.81]{beramono}

\usepackage{graphicx}
\usepackage[table]{xcolor}

\usepackage{fontawesome}

\newcommand{\outOK}[1][green]{\text{\color{#1}\faCheck}}
\newcommand{\outNO}[1][red]{\text{\color{#1}\faClose}}
\newcommand{\outUN}[1][orange]{\text{\color{#1}\faQuestion}}
\definecolor{nocolbg}{rgb}{1,0.8,0.8}

\newcommand{\outcome}[1]{\textnormal{\textsl{#1}}}

\iflong
\else
\fi

\usepackage{tikz}
\usetikzlibrary{shapes,arrows,arrows.meta,positioning,calc,fit,intersections,through}

\definecolor{acol}{HTML}{88a0dc}
\definecolor{bcol}{HTML}{381a61}
\definecolor{ccol}{HTML}{7c4b73}
\definecolor{dcol}{HTML}{ed968c}
\definecolor{ecol}{HTML}{ab3329}
\definecolor{fcol}{HTML}{e78429}
\definecolor{gcol}{HTML}{f9d14a}
\definecolor{hcol}{HTML}{04d17c}

\colorlet{boogiecol}{ccol}
\colorlet{bcccol}{fcol}
\colorlet{pltcol}{ecol}

\usepackage[inline]{enumitem}
\setlist[enumerate]{label=\emph{\roman*})}

\usepackage{ragged2e}

\usepackage{hyperref}
\usepackage{booktabs}
\usepackage{relsize}
\usepackage{changepage}            \usepackage[skip=5pt,font=small,strut=off]{caption}
\usepackage[skip=1pt]{subcaption}  

\graphicspath{{./figs/}}
\DeclareGraphicsExtensions{.pdf,.png}

\usepackage{amsmath,amsfonts}
\usepackage{mathtools}

\usepackage{bussproofs}

\usepackage{array}
\usepackage{xspace}
\usepackage{stmaryrd}

\usepackage{multirow}

\usepackage{pgfkeys,numprint,relsize,refcount}
\xspaceaddexceptions{\%}  \def\NAN{??}              \def\keyfamily{/bcc/}

\DeclareDocumentCommand{\n}{t. t: o m o O{} t| t!}{\begingroup \pgfkeys{/pgf/fpu=true}\IfBooleanTF{#2}{\pgfkeyssetvalue{/tmp/value}{#4}\pgfkeyssetvalue{/tmp/found}{found}}{\pgfkeysifdefined{\keyfamily#4}{\pgfkeyssetvalue{/tmp/value}{\pgfkeysvalueof{\keyfamily#4}}\pgfkeyssetvalue{/tmp/found}{found}}{}}\pgfkeysifdefined{/tmp/found}{\IfNoValueF{#5}{\pgfkeyssetvalue{/tmp/multiplier}{#5}\pgfmathparse{\pgfkeysvalueof{/tmp/multiplier} * \pgfkeysvalueof{/tmp/value}}\pgfkeyslet{/tmp/value}\pgfmathresult }\IfBooleanTF{#1}{\IfBooleanTF{#8}{\spellout{\pgfkeysvalueof{/tmp/value}}}{\pgfkeysvalueof{/tmp/value}}}{\IfNoValueTF{#3}{\pgfmathprintnumber [set thousands separator={\,},int detect,#6]{\pgfkeysvalueof{/tmp/value}}}{{\pgfmathprintnumber [precision=#3,fixed,zerofill,set thousands separator={\,},#6]{\pgfkeysvalueof{/tmp/value}}}}}\IfBooleanT{#7}{{\smaller[1.2]\%}}}{\NAN }\pgfkeys{/pgf/fpu=false}\endgroup }

\pgfkeyssetvalue{/bcc/count/failures}{65332}
\pgfkeyssetvalue{/bcc/perc/failures}{0.021777333333333333}
\pgfkeyssetvalue{/bcc/Term-Counts/well-formed-10/Group}{well-formed}
\pgfkeyssetvalue{/bcc/Term-Counts/well-formed-3/Group}{well-formed}
\pgfkeyssetvalue{/bcc/Term-Counts/well-formed-5/Group}{well-formed}
\pgfkeyssetvalue{/bcc/Term-Counts/well-formed-7/Group}{well-formed}
\pgfkeyssetvalue{/bcc/Term-Counts/well-named-10/Group}{well-named}
\pgfkeyssetvalue{/bcc/Term-Counts/well-named-3/Group}{well-named}
\pgfkeyssetvalue{/bcc/Term-Counts/well-named-5/Group}{well-named}
\pgfkeyssetvalue{/bcc/Term-Counts/well-named-7/Group}{well-named}
\pgfkeyssetvalue{/bcc/Term-Counts/well-typed-10/Group}{well-typed}
\pgfkeyssetvalue{/bcc/Term-Counts/well-typed-3/Group}{well-typed}
\pgfkeyssetvalue{/bcc/Term-Counts/well-typed-5/Group}{well-typed}
\pgfkeyssetvalue{/bcc/Term-Counts/well-typed-7/Group}{well-typed}
\pgfkeyssetvalue{/bcc/Term-Counts/well-formed-10/Size}{10}
\pgfkeyssetvalue{/bcc/Term-Counts/well-formed-3/Size}{3}
\pgfkeyssetvalue{/bcc/Term-Counts/well-formed-5/Size}{5}
\pgfkeyssetvalue{/bcc/Term-Counts/well-formed-7/Size}{7}
\pgfkeyssetvalue{/bcc/Term-Counts/well-named-10/Size}{10}
\pgfkeyssetvalue{/bcc/Term-Counts/well-named-3/Size}{3}
\pgfkeyssetvalue{/bcc/Term-Counts/well-named-5/Size}{5}
\pgfkeyssetvalue{/bcc/Term-Counts/well-named-7/Size}{7}
\pgfkeyssetvalue{/bcc/Term-Counts/well-typed-10/Size}{10}
\pgfkeyssetvalue{/bcc/Term-Counts/well-typed-3/Size}{3}
\pgfkeyssetvalue{/bcc/Term-Counts/well-typed-5/Size}{5}
\pgfkeyssetvalue{/bcc/Term-Counts/well-typed-7/Size}{7}
\pgfkeyssetvalue{/bcc/Term-Counts/well-formed-10/Locals-Minimum}{1}
\pgfkeyssetvalue{/bcc/Term-Counts/well-formed-3/Locals-Minimum}{1}
\pgfkeyssetvalue{/bcc/Term-Counts/well-formed-5/Locals-Minimum}{1}
\pgfkeyssetvalue{/bcc/Term-Counts/well-formed-7/Locals-Minimum}{1}
\pgfkeyssetvalue{/bcc/Term-Counts/well-named-10/Locals-Minimum}{1}
\pgfkeyssetvalue{/bcc/Term-Counts/well-named-3/Locals-Minimum}{1}
\pgfkeyssetvalue{/bcc/Term-Counts/well-named-5/Locals-Minimum}{1}
\pgfkeyssetvalue{/bcc/Term-Counts/well-named-7/Locals-Minimum}{1}
\pgfkeyssetvalue{/bcc/Term-Counts/well-typed-10/Locals-Minimum}{1}
\pgfkeyssetvalue{/bcc/Term-Counts/well-typed-3/Locals-Minimum}{1}
\pgfkeyssetvalue{/bcc/Term-Counts/well-typed-5/Locals-Minimum}{1}
\pgfkeyssetvalue{/bcc/Term-Counts/well-typed-7/Locals-Minimum}{1}
\pgfkeyssetvalue{/bcc/Term-Counts/well-formed-10/Stmt-Count-Minimum}{1}
\pgfkeyssetvalue{/bcc/Term-Counts/well-formed-3/Stmt-Count-Minimum}{1}
\pgfkeyssetvalue{/bcc/Term-Counts/well-formed-5/Stmt-Count-Minimum}{1}
\pgfkeyssetvalue{/bcc/Term-Counts/well-formed-7/Stmt-Count-Minimum}{1}
\pgfkeyssetvalue{/bcc/Term-Counts/well-named-10/Stmt-Count-Minimum}{1}
\pgfkeyssetvalue{/bcc/Term-Counts/well-named-3/Stmt-Count-Minimum}{1}
\pgfkeyssetvalue{/bcc/Term-Counts/well-named-5/Stmt-Count-Minimum}{1}
\pgfkeyssetvalue{/bcc/Term-Counts/well-named-7/Stmt-Count-Minimum}{1}
\pgfkeyssetvalue{/bcc/Term-Counts/well-typed-10/Stmt-Count-Minimum}{1}
\pgfkeyssetvalue{/bcc/Term-Counts/well-typed-3/Stmt-Count-Minimum}{1}
\pgfkeyssetvalue{/bcc/Term-Counts/well-typed-5/Stmt-Count-Minimum}{1}
\pgfkeyssetvalue{/bcc/Term-Counts/well-typed-7/Stmt-Count-Minimum}{1}
\pgfkeyssetvalue{/bcc/Term-Counts/well-formed-10/ARI-Count-Minimum}{0}
\pgfkeyssetvalue{/bcc/Term-Counts/well-formed-3/ARI-Count-Minimum}{0}
\pgfkeyssetvalue{/bcc/Term-Counts/well-formed-5/ARI-Count-Minimum}{0}
\pgfkeyssetvalue{/bcc/Term-Counts/well-formed-7/ARI-Count-Minimum}{0}
\pgfkeyssetvalue{/bcc/Term-Counts/well-named-10/ARI-Count-Minimum}{0}
\pgfkeyssetvalue{/bcc/Term-Counts/well-named-3/ARI-Count-Minimum}{0}
\pgfkeyssetvalue{/bcc/Term-Counts/well-named-5/ARI-Count-Minimum}{0}
\pgfkeyssetvalue{/bcc/Term-Counts/well-named-7/ARI-Count-Minimum}{0}
\pgfkeyssetvalue{/bcc/Term-Counts/well-typed-10/ARI-Count-Minimum}{0}
\pgfkeyssetvalue{/bcc/Term-Counts/well-typed-3/ARI-Count-Minimum}{0}
\pgfkeyssetvalue{/bcc/Term-Counts/well-typed-5/ARI-Count-Minimum}{0}
\pgfkeyssetvalue{/bcc/Term-Counts/well-typed-7/ARI-Count-Minimum}{0}
\pgfkeyssetvalue{/bcc/Term-Counts/well-formed-10/LIT-Count-Minimum}{0}
\pgfkeyssetvalue{/bcc/Term-Counts/well-formed-3/LIT-Count-Minimum}{0}
\pgfkeyssetvalue{/bcc/Term-Counts/well-formed-5/LIT-Count-Minimum}{0}
\pgfkeyssetvalue{/bcc/Term-Counts/well-formed-7/LIT-Count-Minimum}{0}
\pgfkeyssetvalue{/bcc/Term-Counts/well-named-10/LIT-Count-Minimum}{0}
\pgfkeyssetvalue{/bcc/Term-Counts/well-named-3/LIT-Count-Minimum}{0}
\pgfkeyssetvalue{/bcc/Term-Counts/well-named-5/LIT-Count-Minimum}{0}
\pgfkeyssetvalue{/bcc/Term-Counts/well-named-7/LIT-Count-Minimum}{0}
\pgfkeyssetvalue{/bcc/Term-Counts/well-typed-10/LIT-Count-Minimum}{0}
\pgfkeyssetvalue{/bcc/Term-Counts/well-typed-3/LIT-Count-Minimum}{0}
\pgfkeyssetvalue{/bcc/Term-Counts/well-typed-5/LIT-Count-Minimum}{0}
\pgfkeyssetvalue{/bcc/Term-Counts/well-typed-7/LIT-Count-Minimum}{0}
\pgfkeyssetvalue{/bcc/Term-Counts/well-formed-10/LOG-Count-Minimum}{0}
\pgfkeyssetvalue{/bcc/Term-Counts/well-formed-3/LOG-Count-Minimum}{0}
\pgfkeyssetvalue{/bcc/Term-Counts/well-formed-5/LOG-Count-Minimum}{0}
\pgfkeyssetvalue{/bcc/Term-Counts/well-formed-7/LOG-Count-Minimum}{0}
\pgfkeyssetvalue{/bcc/Term-Counts/well-named-10/LOG-Count-Minimum}{0}
\pgfkeyssetvalue{/bcc/Term-Counts/well-named-3/LOG-Count-Minimum}{0}
\pgfkeyssetvalue{/bcc/Term-Counts/well-named-5/LOG-Count-Minimum}{0}
\pgfkeyssetvalue{/bcc/Term-Counts/well-named-7/LOG-Count-Minimum}{0}
\pgfkeyssetvalue{/bcc/Term-Counts/well-typed-10/LOG-Count-Minimum}{0}
\pgfkeyssetvalue{/bcc/Term-Counts/well-typed-3/LOG-Count-Minimum}{0}
\pgfkeyssetvalue{/bcc/Term-Counts/well-typed-5/LOG-Count-Minimum}{0}
\pgfkeyssetvalue{/bcc/Term-Counts/well-typed-7/LOG-Count-Minimum}{0}
\pgfkeyssetvalue{/bcc/Term-Counts/well-formed-10/CMP-Count-Minimum}{0}
\pgfkeyssetvalue{/bcc/Term-Counts/well-formed-3/CMP-Count-Minimum}{0}
\pgfkeyssetvalue{/bcc/Term-Counts/well-formed-5/CMP-Count-Minimum}{0}
\pgfkeyssetvalue{/bcc/Term-Counts/well-formed-7/CMP-Count-Minimum}{0}
\pgfkeyssetvalue{/bcc/Term-Counts/well-named-10/CMP-Count-Minimum}{0}
\pgfkeyssetvalue{/bcc/Term-Counts/well-named-3/CMP-Count-Minimum}{0}
\pgfkeyssetvalue{/bcc/Term-Counts/well-named-5/CMP-Count-Minimum}{0}
\pgfkeyssetvalue{/bcc/Term-Counts/well-named-7/CMP-Count-Minimum}{0}
\pgfkeyssetvalue{/bcc/Term-Counts/well-typed-10/CMP-Count-Minimum}{0}
\pgfkeyssetvalue{/bcc/Term-Counts/well-typed-3/CMP-Count-Minimum}{0}
\pgfkeyssetvalue{/bcc/Term-Counts/well-typed-5/CMP-Count-Minimum}{0}
\pgfkeyssetvalue{/bcc/Term-Counts/well-typed-7/CMP-Count-Minimum}{0}
\pgfkeyssetvalue{/bcc/Term-Counts/well-formed-10/VAR-Count-Minimum}{0}
\pgfkeyssetvalue{/bcc/Term-Counts/well-formed-3/VAR-Count-Minimum}{0}
\pgfkeyssetvalue{/bcc/Term-Counts/well-formed-5/VAR-Count-Minimum}{0}
\pgfkeyssetvalue{/bcc/Term-Counts/well-formed-7/VAR-Count-Minimum}{0}
\pgfkeyssetvalue{/bcc/Term-Counts/well-named-10/VAR-Count-Minimum}{0}
\pgfkeyssetvalue{/bcc/Term-Counts/well-named-3/VAR-Count-Minimum}{0}
\pgfkeyssetvalue{/bcc/Term-Counts/well-named-5/VAR-Count-Minimum}{0}
\pgfkeyssetvalue{/bcc/Term-Counts/well-named-7/VAR-Count-Minimum}{0}
\pgfkeyssetvalue{/bcc/Term-Counts/well-typed-10/VAR-Count-Minimum}{0}
\pgfkeyssetvalue{/bcc/Term-Counts/well-typed-3/VAR-Count-Minimum}{0}
\pgfkeyssetvalue{/bcc/Term-Counts/well-typed-5/VAR-Count-Minimum}{0}
\pgfkeyssetvalue{/bcc/Term-Counts/well-typed-7/VAR-Count-Minimum}{0}
\pgfkeyssetvalue{/bcc/Term-Counts/well-formed-10/Locals-Median}{2.0}
\pgfkeyssetvalue{/bcc/Term-Counts/well-formed-3/Locals-Median}{1.0}
\pgfkeyssetvalue{/bcc/Term-Counts/well-formed-5/Locals-Median}{1.0}
\pgfkeyssetvalue{/bcc/Term-Counts/well-formed-7/Locals-Median}{1.0}
\pgfkeyssetvalue{/bcc/Term-Counts/well-named-10/Locals-Median}{1.0}
\pgfkeyssetvalue{/bcc/Term-Counts/well-named-3/Locals-Median}{1.0}
\pgfkeyssetvalue{/bcc/Term-Counts/well-named-5/Locals-Median}{1.0}
\pgfkeyssetvalue{/bcc/Term-Counts/well-named-7/Locals-Median}{1.0}
\pgfkeyssetvalue{/bcc/Term-Counts/well-typed-10/Locals-Median}{2.0}
\pgfkeyssetvalue{/bcc/Term-Counts/well-typed-3/Locals-Median}{1.0}
\pgfkeyssetvalue{/bcc/Term-Counts/well-typed-5/Locals-Median}{1.0}
\pgfkeyssetvalue{/bcc/Term-Counts/well-typed-7/Locals-Median}{1.0}
\pgfkeyssetvalue{/bcc/Term-Counts/well-formed-10/Stmt-Count-Median}{100.0}
\pgfkeyssetvalue{/bcc/Term-Counts/well-formed-3/Stmt-Count-Median}{9.0}
\pgfkeyssetvalue{/bcc/Term-Counts/well-formed-5/Stmt-Count-Median}{20.0}
\pgfkeyssetvalue{/bcc/Term-Counts/well-formed-7/Stmt-Count-Median}{40.0}
\pgfkeyssetvalue{/bcc/Term-Counts/well-named-10/Stmt-Count-Median}{102.0}
\pgfkeyssetvalue{/bcc/Term-Counts/well-named-3/Stmt-Count-Median}{10.0}
\pgfkeyssetvalue{/bcc/Term-Counts/well-named-5/Stmt-Count-Median}{21.0}
\pgfkeyssetvalue{/bcc/Term-Counts/well-named-7/Stmt-Count-Median}{41.0}
\pgfkeyssetvalue{/bcc/Term-Counts/well-typed-10/Stmt-Count-Median}{103.0}
\pgfkeyssetvalue{/bcc/Term-Counts/well-typed-3/Stmt-Count-Median}{10.0}
\pgfkeyssetvalue{/bcc/Term-Counts/well-typed-5/Stmt-Count-Median}{21.0}
\pgfkeyssetvalue{/bcc/Term-Counts/well-typed-7/Stmt-Count-Median}{41.0}
\pgfkeyssetvalue{/bcc/Term-Counts/well-formed-10/ARI-Count-Median}{16.0}
\pgfkeyssetvalue{/bcc/Term-Counts/well-formed-3/ARI-Count-Median}{1.0}
\pgfkeyssetvalue{/bcc/Term-Counts/well-formed-5/ARI-Count-Median}{2.0}
\pgfkeyssetvalue{/bcc/Term-Counts/well-formed-7/ARI-Count-Median}{5.0}
\pgfkeyssetvalue{/bcc/Term-Counts/well-named-10/ARI-Count-Median}{14.0}
\pgfkeyssetvalue{/bcc/Term-Counts/well-named-3/ARI-Count-Median}{0.0}
\pgfkeyssetvalue{/bcc/Term-Counts/well-named-5/ARI-Count-Median}{2.0}
\pgfkeyssetvalue{/bcc/Term-Counts/well-named-7/ARI-Count-Median}{4.0}
\pgfkeyssetvalue{/bcc/Term-Counts/well-typed-10/ARI-Count-Median}{37.0}
\pgfkeyssetvalue{/bcc/Term-Counts/well-typed-3/ARI-Count-Median}{0.0}
\pgfkeyssetvalue{/bcc/Term-Counts/well-typed-5/ARI-Count-Median}{3.0}
\pgfkeyssetvalue{/bcc/Term-Counts/well-typed-7/ARI-Count-Median}{9.0}
\pgfkeyssetvalue{/bcc/Term-Counts/well-formed-10/LIT-Count-Median}{79.0}
\pgfkeyssetvalue{/bcc/Term-Counts/well-formed-3/LIT-Count-Median}{7.0}
\pgfkeyssetvalue{/bcc/Term-Counts/well-formed-5/LIT-Count-Median}{15.0}
\pgfkeyssetvalue{/bcc/Term-Counts/well-formed-7/LIT-Count-Median}{31.0}
\pgfkeyssetvalue{/bcc/Term-Counts/well-named-10/LIT-Count-Median}{93.0}
\pgfkeyssetvalue{/bcc/Term-Counts/well-named-3/LIT-Count-Median}{10.0}
\pgfkeyssetvalue{/bcc/Term-Counts/well-named-5/LIT-Count-Median}{20.0}
\pgfkeyssetvalue{/bcc/Term-Counts/well-named-7/LIT-Count-Median}{38.0}
\pgfkeyssetvalue{/bcc/Term-Counts/well-typed-10/LIT-Count-Median}{106.0}
\pgfkeyssetvalue{/bcc/Term-Counts/well-typed-3/LIT-Count-Median}{8.0}
\pgfkeyssetvalue{/bcc/Term-Counts/well-typed-5/LIT-Count-Median}{18.0}
\pgfkeyssetvalue{/bcc/Term-Counts/well-typed-7/LIT-Count-Median}{38.0}
\pgfkeyssetvalue{/bcc/Term-Counts/well-formed-10/LOG-Count-Median}{54.0}
\pgfkeyssetvalue{/bcc/Term-Counts/well-formed-3/LOG-Count-Median}{4.0}
\pgfkeyssetvalue{/bcc/Term-Counts/well-formed-5/LOG-Count-Median}{10.0}
\pgfkeyssetvalue{/bcc/Term-Counts/well-formed-7/LOG-Count-Median}{20.0}
\pgfkeyssetvalue{/bcc/Term-Counts/well-named-10/LOG-Count-Median}{60.0}
\pgfkeyssetvalue{/bcc/Term-Counts/well-named-3/LOG-Count-Median}{5.0}
\pgfkeyssetvalue{/bcc/Term-Counts/well-named-5/LOG-Count-Median}{12.0}
\pgfkeyssetvalue{/bcc/Term-Counts/well-named-7/LOG-Count-Median}{23.0}
\pgfkeyssetvalue{/bcc/Term-Counts/well-typed-10/LOG-Count-Median}{94.0}
\pgfkeyssetvalue{/bcc/Term-Counts/well-typed-3/LOG-Count-Median}{7.0}
\pgfkeyssetvalue{/bcc/Term-Counts/well-typed-5/LOG-Count-Median}{17.0}
\pgfkeyssetvalue{/bcc/Term-Counts/well-typed-7/LOG-Count-Median}{35.0}
\pgfkeyssetvalue{/bcc/Term-Counts/well-formed-10/CMP-Count-Median}{8.0}
\pgfkeyssetvalue{/bcc/Term-Counts/well-formed-3/CMP-Count-Median}{0.0}
\pgfkeyssetvalue{/bcc/Term-Counts/well-formed-5/CMP-Count-Median}{1.0}
\pgfkeyssetvalue{/bcc/Term-Counts/well-formed-7/CMP-Count-Median}{2.0}
\pgfkeyssetvalue{/bcc/Term-Counts/well-named-10/CMP-Count-Median}{7.0}
\pgfkeyssetvalue{/bcc/Term-Counts/well-named-3/CMP-Count-Median}{0.0}
\pgfkeyssetvalue{/bcc/Term-Counts/well-named-5/CMP-Count-Median}{1.0}
\pgfkeyssetvalue{/bcc/Term-Counts/well-named-7/CMP-Count-Median}{2.0}
\pgfkeyssetvalue{/bcc/Term-Counts/well-typed-10/CMP-Count-Median}{24.0}
\pgfkeyssetvalue{/bcc/Term-Counts/well-typed-3/CMP-Count-Median}{1.0}
\pgfkeyssetvalue{/bcc/Term-Counts/well-typed-5/CMP-Count-Median}{3.0}
\pgfkeyssetvalue{/bcc/Term-Counts/well-typed-7/CMP-Count-Median}{7.0}
\pgfkeyssetvalue{/bcc/Term-Counts/well-formed-10/VAR-Count-Median}{61.0}
\pgfkeyssetvalue{/bcc/Term-Counts/well-formed-3/VAR-Count-Median}{5.0}
\pgfkeyssetvalue{/bcc/Term-Counts/well-formed-5/VAR-Count-Median}{11.0}
\pgfkeyssetvalue{/bcc/Term-Counts/well-formed-7/VAR-Count-Median}{23.0}
\pgfkeyssetvalue{/bcc/Term-Counts/well-named-10/VAR-Count-Median}{45.0}
\pgfkeyssetvalue{/bcc/Term-Counts/well-named-3/VAR-Count-Median}{2.0}
\pgfkeyssetvalue{/bcc/Term-Counts/well-named-5/VAR-Count-Median}{6.0}
\pgfkeyssetvalue{/bcc/Term-Counts/well-named-7/VAR-Count-Median}{15.0}
\pgfkeyssetvalue{/bcc/Term-Counts/well-typed-10/VAR-Count-Median}{75.0}
\pgfkeyssetvalue{/bcc/Term-Counts/well-typed-3/VAR-Count-Median}{5.0}
\pgfkeyssetvalue{/bcc/Term-Counts/well-typed-5/VAR-Count-Median}{12.0}
\pgfkeyssetvalue{/bcc/Term-Counts/well-typed-7/VAR-Count-Median}{26.0}
\pgfkeyssetvalue{/bcc/Term-Counts/well-formed-10/Locals-Maximum}{10}
\pgfkeyssetvalue{/bcc/Term-Counts/well-formed-3/Locals-Maximum}{3}
\pgfkeyssetvalue{/bcc/Term-Counts/well-formed-5/Locals-Maximum}{5}
\pgfkeyssetvalue{/bcc/Term-Counts/well-formed-7/Locals-Maximum}{7}
\pgfkeyssetvalue{/bcc/Term-Counts/well-named-10/Locals-Maximum}{10}
\pgfkeyssetvalue{/bcc/Term-Counts/well-named-3/Locals-Maximum}{3}
\pgfkeyssetvalue{/bcc/Term-Counts/well-named-5/Locals-Maximum}{5}
\pgfkeyssetvalue{/bcc/Term-Counts/well-named-7/Locals-Maximum}{7}
\pgfkeyssetvalue{/bcc/Term-Counts/well-typed-10/Locals-Maximum}{10}
\pgfkeyssetvalue{/bcc/Term-Counts/well-typed-3/Locals-Maximum}{3}
\pgfkeyssetvalue{/bcc/Term-Counts/well-typed-5/Locals-Maximum}{5}
\pgfkeyssetvalue{/bcc/Term-Counts/well-typed-7/Locals-Maximum}{7}
\pgfkeyssetvalue{/bcc/Term-Counts/well-formed-10/Stmt-Count-Maximum}{637}
\pgfkeyssetvalue{/bcc/Term-Counts/well-formed-3/Stmt-Count-Maximum}{18}
\pgfkeyssetvalue{/bcc/Term-Counts/well-formed-5/Stmt-Count-Maximum}{65}
\pgfkeyssetvalue{/bcc/Term-Counts/well-formed-7/Stmt-Count-Maximum}{176}
\pgfkeyssetvalue{/bcc/Term-Counts/well-named-10/Stmt-Count-Maximum}{491}
\pgfkeyssetvalue{/bcc/Term-Counts/well-named-3/Stmt-Count-Maximum}{18}
\pgfkeyssetvalue{/bcc/Term-Counts/well-named-5/Stmt-Count-Maximum}{61}
\pgfkeyssetvalue{/bcc/Term-Counts/well-named-7/Stmt-Count-Maximum}{130}
\pgfkeyssetvalue{/bcc/Term-Counts/well-typed-10/Stmt-Count-Maximum}{481}
\pgfkeyssetvalue{/bcc/Term-Counts/well-typed-3/Stmt-Count-Maximum}{18}
\pgfkeyssetvalue{/bcc/Term-Counts/well-typed-5/Stmt-Count-Maximum}{54}
\pgfkeyssetvalue{/bcc/Term-Counts/well-typed-7/Stmt-Count-Maximum}{139}
\pgfkeyssetvalue{/bcc/Term-Counts/well-formed-10/ARI-Count-Maximum}{126}
\pgfkeyssetvalue{/bcc/Term-Counts/well-formed-3/ARI-Count-Maximum}{5}
\pgfkeyssetvalue{/bcc/Term-Counts/well-formed-5/ARI-Count-Maximum}{14}
\pgfkeyssetvalue{/bcc/Term-Counts/well-formed-7/ARI-Count-Maximum}{39}
\pgfkeyssetvalue{/bcc/Term-Counts/well-named-10/ARI-Count-Maximum}{108}
\pgfkeyssetvalue{/bcc/Term-Counts/well-named-3/ARI-Count-Maximum}{5}
\pgfkeyssetvalue{/bcc/Term-Counts/well-named-5/ARI-Count-Maximum}{15}
\pgfkeyssetvalue{/bcc/Term-Counts/well-named-7/ARI-Count-Maximum}{31}
\pgfkeyssetvalue{/bcc/Term-Counts/well-typed-10/ARI-Count-Maximum}{265}
\pgfkeyssetvalue{/bcc/Term-Counts/well-typed-3/ARI-Count-Maximum}{4}
\pgfkeyssetvalue{/bcc/Term-Counts/well-typed-5/ARI-Count-Maximum}{17}
\pgfkeyssetvalue{/bcc/Term-Counts/well-typed-7/ARI-Count-Maximum}{48}
\pgfkeyssetvalue{/bcc/Term-Counts/well-formed-10/LIT-Count-Maximum}{517}
\pgfkeyssetvalue{/bcc/Term-Counts/well-formed-3/LIT-Count-Maximum}{16}
\pgfkeyssetvalue{/bcc/Term-Counts/well-formed-5/LIT-Count-Maximum}{51}
\pgfkeyssetvalue{/bcc/Term-Counts/well-formed-7/LIT-Count-Maximum}{132}
\pgfkeyssetvalue{/bcc/Term-Counts/well-named-10/LIT-Count-Maximum}{471}
\pgfkeyssetvalue{/bcc/Term-Counts/well-named-3/LIT-Count-Maximum}{20}
\pgfkeyssetvalue{/bcc/Term-Counts/well-named-5/LIT-Count-Maximum}{57}
\pgfkeyssetvalue{/bcc/Term-Counts/well-named-7/LIT-Count-Maximum}{125}
\pgfkeyssetvalue{/bcc/Term-Counts/well-typed-10/LIT-Count-Maximum}{584}
\pgfkeyssetvalue{/bcc/Term-Counts/well-typed-3/LIT-Count-Maximum}{20}
\pgfkeyssetvalue{/bcc/Term-Counts/well-typed-5/LIT-Count-Maximum}{58}
\pgfkeyssetvalue{/bcc/Term-Counts/well-typed-7/LIT-Count-Maximum}{170}
\pgfkeyssetvalue{/bcc/Term-Counts/well-formed-10/LOG-Count-Maximum}{379}
\pgfkeyssetvalue{/bcc/Term-Counts/well-formed-3/LOG-Count-Maximum}{14}
\pgfkeyssetvalue{/bcc/Term-Counts/well-formed-5/LOG-Count-Maximum}{37}
\pgfkeyssetvalue{/bcc/Term-Counts/well-formed-7/LOG-Count-Maximum}{110}
\pgfkeyssetvalue{/bcc/Term-Counts/well-named-10/LOG-Count-Maximum}{326}
\pgfkeyssetvalue{/bcc/Term-Counts/well-named-3/LOG-Count-Maximum}{17}
\pgfkeyssetvalue{/bcc/Term-Counts/well-named-5/LOG-Count-Maximum}{41}
\pgfkeyssetvalue{/bcc/Term-Counts/well-named-7/LOG-Count-Maximum}{106}
\pgfkeyssetvalue{/bcc/Term-Counts/well-typed-10/LOG-Count-Maximum}{553}
\pgfkeyssetvalue{/bcc/Term-Counts/well-typed-3/LOG-Count-Maximum}{27}
\pgfkeyssetvalue{/bcc/Term-Counts/well-typed-5/LOG-Count-Maximum}{65}
\pgfkeyssetvalue{/bcc/Term-Counts/well-typed-7/LOG-Count-Maximum}{187}
\pgfkeyssetvalue{/bcc/Term-Counts/well-formed-10/CMP-Count-Maximum}{78}
\pgfkeyssetvalue{/bcc/Term-Counts/well-formed-3/CMP-Count-Maximum}{5}
\pgfkeyssetvalue{/bcc/Term-Counts/well-formed-5/CMP-Count-Maximum}{11}
\pgfkeyssetvalue{/bcc/Term-Counts/well-formed-7/CMP-Count-Maximum}{22}
\pgfkeyssetvalue{/bcc/Term-Counts/well-named-10/CMP-Count-Maximum}{66}
\pgfkeyssetvalue{/bcc/Term-Counts/well-named-3/CMP-Count-Maximum}{5}
\pgfkeyssetvalue{/bcc/Term-Counts/well-named-5/CMP-Count-Maximum}{11}
\pgfkeyssetvalue{/bcc/Term-Counts/well-named-7/CMP-Count-Maximum}{19}
\pgfkeyssetvalue{/bcc/Term-Counts/well-typed-10/CMP-Count-Maximum}{157}
\pgfkeyssetvalue{/bcc/Term-Counts/well-typed-3/CMP-Count-Maximum}{6}
\pgfkeyssetvalue{/bcc/Term-Counts/well-typed-5/CMP-Count-Maximum}{16}
\pgfkeyssetvalue{/bcc/Term-Counts/well-typed-7/CMP-Count-Maximum}{49}
\pgfkeyssetvalue{/bcc/Term-Counts/well-formed-10/VAR-Count-Maximum}{402}
\pgfkeyssetvalue{/bcc/Term-Counts/well-formed-3/VAR-Count-Maximum}{15}
\pgfkeyssetvalue{/bcc/Term-Counts/well-formed-5/VAR-Count-Maximum}{42}
\pgfkeyssetvalue{/bcc/Term-Counts/well-formed-7/VAR-Count-Maximum}{110}
\pgfkeyssetvalue{/bcc/Term-Counts/well-named-10/VAR-Count-Maximum}{274}
\pgfkeyssetvalue{/bcc/Term-Counts/well-named-3/VAR-Count-Maximum}{8}
\pgfkeyssetvalue{/bcc/Term-Counts/well-named-5/VAR-Count-Maximum}{26}
\pgfkeyssetvalue{/bcc/Term-Counts/well-named-7/VAR-Count-Maximum}{64}
\pgfkeyssetvalue{/bcc/Term-Counts/well-typed-10/VAR-Count-Maximum}{367}
\pgfkeyssetvalue{/bcc/Term-Counts/well-typed-3/VAR-Count-Maximum}{17}
\pgfkeyssetvalue{/bcc/Term-Counts/well-typed-5/VAR-Count-Maximum}{50}
\pgfkeyssetvalue{/bcc/Term-Counts/well-typed-7/VAR-Count-Maximum}{104}

\pgfkeyssetvalue{/bcc/Execution-Outcomes/name-error/Execution-Outcome}{name-error}
\pgfkeyssetvalue{/bcc/Execution-Outcomes/type-error/Execution-Outcome}{type-error}
\pgfkeyssetvalue{/bcc/Execution-Outcomes/success/Execution-Outcome}{success}
\pgfkeyssetvalue{/bcc/Execution-Outcomes/failure/Execution-Outcome}{failure}
\pgfkeyssetvalue{/bcc/Execution-Outcomes/loop/Execution-Outcome}{loop}
\pgfkeyssetvalue{/bcc/Execution-Outcomes/timeout/Execution-Outcome}{timeout}
\pgfkeyssetvalue{/bcc/Execution-Outcomes/name-error/Count}{999301}
\pgfkeyssetvalue{/bcc/Execution-Outcomes/type-error/Count}{1000437}
\pgfkeyssetvalue{/bcc/Execution-Outcomes/success/Count}{7330}
\pgfkeyssetvalue{/bcc/Execution-Outcomes/failure/Count}{667408}
\pgfkeyssetvalue{/bcc/Execution-Outcomes/loop/Count}{323821}
\pgfkeyssetvalue{/bcc/Execution-Outcomes/timeout/Count}{1703}
\pgfkeyssetvalue{/bcc/Execution-Outcomes/name-error/Percentage}{33.31003333333334}
\pgfkeyssetvalue{/bcc/Execution-Outcomes/type-error/Percentage}{33.3479}
\pgfkeyssetvalue{/bcc/Execution-Outcomes/success/Percentage}{0.24433333333333335}
\pgfkeyssetvalue{/bcc/Execution-Outcomes/failure/Percentage}{22.24693333333333}
\pgfkeyssetvalue{/bcc/Execution-Outcomes/loop/Percentage}{10.794033333333333}
\pgfkeyssetvalue{/bcc/Execution-Outcomes/timeout/Percentage}{0.05676666666666666}

\pgfkeyssetvalue{/bcc/Execution-Outcomes/well-formed-name-error/Group}{well-formed}
\pgfkeyssetvalue{/bcc/Execution-Outcomes/well-formed-type-error/Group}{well-formed}
\pgfkeyssetvalue{/bcc/Execution-Outcomes/well-formed-success/Group}{well-formed}
\pgfkeyssetvalue{/bcc/Execution-Outcomes/well-formed-failure/Group}{well-formed}
\pgfkeyssetvalue{/bcc/Execution-Outcomes/well-formed-loop/Group}{well-formed}
\pgfkeyssetvalue{/bcc/Execution-Outcomes/well-named-type-error/Group}{well-named}
\pgfkeyssetvalue{/bcc/Execution-Outcomes/well-named-success/Group}{well-named}
\pgfkeyssetvalue{/bcc/Execution-Outcomes/well-named-failure/Group}{well-named}
\pgfkeyssetvalue{/bcc/Execution-Outcomes/well-named-loop/Group}{well-named}
\pgfkeyssetvalue{/bcc/Execution-Outcomes/well-typed-failure/Group}{well-typed}
\pgfkeyssetvalue{/bcc/Execution-Outcomes/well-typed-loop/Group}{well-typed}
\pgfkeyssetvalue{/bcc/Execution-Outcomes/well-typed-timeout/Group}{well-typed}
\pgfkeyssetvalue{/bcc/Execution-Outcomes/well-typed-success/Group}{well-typed}
\pgfkeyssetvalue{/bcc/Execution-Outcomes/well-formed-name-error/Execution-Outcome}{name-error}
\pgfkeyssetvalue{/bcc/Execution-Outcomes/well-formed-type-error/Execution-Outcome}{type-error}
\pgfkeyssetvalue{/bcc/Execution-Outcomes/well-formed-success/Execution-Outcome}{success}
\pgfkeyssetvalue{/bcc/Execution-Outcomes/well-formed-failure/Execution-Outcome}{failure}
\pgfkeyssetvalue{/bcc/Execution-Outcomes/well-formed-loop/Execution-Outcome}{loop}
\pgfkeyssetvalue{/bcc/Execution-Outcomes/well-named-type-error/Execution-Outcome}{type-error}
\pgfkeyssetvalue{/bcc/Execution-Outcomes/well-named-success/Execution-Outcome}{success}
\pgfkeyssetvalue{/bcc/Execution-Outcomes/well-named-failure/Execution-Outcome}{failure}
\pgfkeyssetvalue{/bcc/Execution-Outcomes/well-named-loop/Execution-Outcome}{loop}
\pgfkeyssetvalue{/bcc/Execution-Outcomes/well-typed-failure/Execution-Outcome}{failure}
\pgfkeyssetvalue{/bcc/Execution-Outcomes/well-typed-loop/Execution-Outcome}{loop}
\pgfkeyssetvalue{/bcc/Execution-Outcomes/well-typed-timeout/Execution-Outcome}{timeout}
\pgfkeyssetvalue{/bcc/Execution-Outcomes/well-typed-success/Execution-Outcome}{success}
\pgfkeyssetvalue{/bcc/Execution-Outcomes/well-formed-name-error/Count}{999301}
\pgfkeyssetvalue{/bcc/Execution-Outcomes/well-formed-type-error/Count}{663}
\pgfkeyssetvalue{/bcc/Execution-Outcomes/well-formed-success/Count}{24}
\pgfkeyssetvalue{/bcc/Execution-Outcomes/well-formed-failure/Count}{10}
\pgfkeyssetvalue{/bcc/Execution-Outcomes/well-formed-loop/Count}{2}
\pgfkeyssetvalue{/bcc/Execution-Outcomes/well-named-type-error/Count}{999774}
\pgfkeyssetvalue{/bcc/Execution-Outcomes/well-named-success/Count}{144}
\pgfkeyssetvalue{/bcc/Execution-Outcomes/well-named-failure/Count}{58}
\pgfkeyssetvalue{/bcc/Execution-Outcomes/well-named-loop/Count}{24}
\pgfkeyssetvalue{/bcc/Execution-Outcomes/well-typed-failure/Count}{667340}
\pgfkeyssetvalue{/bcc/Execution-Outcomes/well-typed-loop/Count}{323795}
\pgfkeyssetvalue{/bcc/Execution-Outcomes/well-typed-timeout/Count}{1703}
\pgfkeyssetvalue{/bcc/Execution-Outcomes/well-typed-success/Count}{7162}
\pgfkeyssetvalue{/bcc/Execution-Outcomes/well-formed-name-error/Percentage}{99.9301}
\pgfkeyssetvalue{/bcc/Execution-Outcomes/well-formed-type-error/Percentage}{0.0663}
\pgfkeyssetvalue{/bcc/Execution-Outcomes/well-formed-success/Percentage}{0.0024000000000000002}
\pgfkeyssetvalue{/bcc/Execution-Outcomes/well-formed-failure/Percentage}{0.001}
\pgfkeyssetvalue{/bcc/Execution-Outcomes/well-formed-loop/Percentage}{0.00019999999999999998}
\pgfkeyssetvalue{/bcc/Execution-Outcomes/well-named-type-error/Percentage}{99.9774}
\pgfkeyssetvalue{/bcc/Execution-Outcomes/well-named-success/Percentage}{0.0144}
\pgfkeyssetvalue{/bcc/Execution-Outcomes/well-named-failure/Percentage}{0.0058}
\pgfkeyssetvalue{/bcc/Execution-Outcomes/well-named-loop/Percentage}{0.0024000000000000002}
\pgfkeyssetvalue{/bcc/Execution-Outcomes/well-typed-failure/Percentage}{66.73400000000001}
\pgfkeyssetvalue{/bcc/Execution-Outcomes/well-typed-loop/Percentage}{32.3795}
\pgfkeyssetvalue{/bcc/Execution-Outcomes/well-typed-timeout/Percentage}{0.1703}
\pgfkeyssetvalue{/bcc/Execution-Outcomes/well-typed-success/Percentage}{0.7162}

\pgfkeyssetvalue{/bcc/Execution-Outcomes/well-formed-10-name-error/Group}{well-formed}
\pgfkeyssetvalue{/bcc/Execution-Outcomes/well-formed-10-type-error/Group}{well-formed}
\pgfkeyssetvalue{/bcc/Execution-Outcomes/well-formed-10-success/Group}{well-formed}
\pgfkeyssetvalue{/bcc/Execution-Outcomes/well-formed-10-failure/Group}{well-formed}
\pgfkeyssetvalue{/bcc/Execution-Outcomes/well-formed-10-loop/Group}{well-formed}
\pgfkeyssetvalue{/bcc/Execution-Outcomes/well-formed-3-name-error/Group}{well-formed}
\pgfkeyssetvalue{/bcc/Execution-Outcomes/well-formed-3-type-error/Group}{well-formed}
\pgfkeyssetvalue{/bcc/Execution-Outcomes/well-formed-3-success/Group}{well-formed}
\pgfkeyssetvalue{/bcc/Execution-Outcomes/well-formed-5-name-error/Group}{well-formed}
\pgfkeyssetvalue{/bcc/Execution-Outcomes/well-formed-5-type-error/Group}{well-formed}
\pgfkeyssetvalue{/bcc/Execution-Outcomes/well-formed-5-failure/Group}{well-formed}
\pgfkeyssetvalue{/bcc/Execution-Outcomes/well-formed-5-success/Group}{well-formed}
\pgfkeyssetvalue{/bcc/Execution-Outcomes/well-formed-7-name-error/Group}{well-formed}
\pgfkeyssetvalue{/bcc/Execution-Outcomes/well-formed-7-success/Group}{well-formed}
\pgfkeyssetvalue{/bcc/Execution-Outcomes/well-formed-7-type-error/Group}{well-formed}
\pgfkeyssetvalue{/bcc/Execution-Outcomes/well-formed-7-failure/Group}{well-formed}
\pgfkeyssetvalue{/bcc/Execution-Outcomes/well-formed-7-loop/Group}{well-formed}
\pgfkeyssetvalue{/bcc/Execution-Outcomes/well-named-10-type-error/Group}{well-named}
\pgfkeyssetvalue{/bcc/Execution-Outcomes/well-named-10-success/Group}{well-named}
\pgfkeyssetvalue{/bcc/Execution-Outcomes/well-named-10-failure/Group}{well-named}
\pgfkeyssetvalue{/bcc/Execution-Outcomes/well-named-10-loop/Group}{well-named}
\pgfkeyssetvalue{/bcc/Execution-Outcomes/well-named-3-type-error/Group}{well-named}
\pgfkeyssetvalue{/bcc/Execution-Outcomes/well-named-3-failure/Group}{well-named}
\pgfkeyssetvalue{/bcc/Execution-Outcomes/well-named-3-loop/Group}{well-named}
\pgfkeyssetvalue{/bcc/Execution-Outcomes/well-named-3-success/Group}{well-named}
\pgfkeyssetvalue{/bcc/Execution-Outcomes/well-named-5-type-error/Group}{well-named}
\pgfkeyssetvalue{/bcc/Execution-Outcomes/well-named-5-success/Group}{well-named}
\pgfkeyssetvalue{/bcc/Execution-Outcomes/well-named-5-failure/Group}{well-named}
\pgfkeyssetvalue{/bcc/Execution-Outcomes/well-named-5-loop/Group}{well-named}
\pgfkeyssetvalue{/bcc/Execution-Outcomes/well-named-7-type-error/Group}{well-named}
\pgfkeyssetvalue{/bcc/Execution-Outcomes/well-named-7-success/Group}{well-named}
\pgfkeyssetvalue{/bcc/Execution-Outcomes/well-named-7-failure/Group}{well-named}
\pgfkeyssetvalue{/bcc/Execution-Outcomes/well-named-7-loop/Group}{well-named}
\pgfkeyssetvalue{/bcc/Execution-Outcomes/well-typed-10-failure/Group}{well-typed}
\pgfkeyssetvalue{/bcc/Execution-Outcomes/well-typed-10-loop/Group}{well-typed}
\pgfkeyssetvalue{/bcc/Execution-Outcomes/well-typed-10-timeout/Group}{well-typed}
\pgfkeyssetvalue{/bcc/Execution-Outcomes/well-typed-10-success/Group}{well-typed}
\pgfkeyssetvalue{/bcc/Execution-Outcomes/well-typed-3-failure/Group}{well-typed}
\pgfkeyssetvalue{/bcc/Execution-Outcomes/well-typed-3-loop/Group}{well-typed}
\pgfkeyssetvalue{/bcc/Execution-Outcomes/well-typed-3-success/Group}{well-typed}
\pgfkeyssetvalue{/bcc/Execution-Outcomes/well-typed-3-timeout/Group}{well-typed}
\pgfkeyssetvalue{/bcc/Execution-Outcomes/well-typed-5-failure/Group}{well-typed}
\pgfkeyssetvalue{/bcc/Execution-Outcomes/well-typed-5-loop/Group}{well-typed}
\pgfkeyssetvalue{/bcc/Execution-Outcomes/well-typed-5-success/Group}{well-typed}
\pgfkeyssetvalue{/bcc/Execution-Outcomes/well-typed-5-timeout/Group}{well-typed}
\pgfkeyssetvalue{/bcc/Execution-Outcomes/well-typed-7-loop/Group}{well-typed}
\pgfkeyssetvalue{/bcc/Execution-Outcomes/well-typed-7-failure/Group}{well-typed}
\pgfkeyssetvalue{/bcc/Execution-Outcomes/well-typed-7-success/Group}{well-typed}
\pgfkeyssetvalue{/bcc/Execution-Outcomes/well-typed-7-timeout/Group}{well-typed}
\pgfkeyssetvalue{/bcc/Execution-Outcomes/well-formed-10-name-error/Size}{10}
\pgfkeyssetvalue{/bcc/Execution-Outcomes/well-formed-10-type-error/Size}{10}
\pgfkeyssetvalue{/bcc/Execution-Outcomes/well-formed-10-success/Size}{10}
\pgfkeyssetvalue{/bcc/Execution-Outcomes/well-formed-10-failure/Size}{10}
\pgfkeyssetvalue{/bcc/Execution-Outcomes/well-formed-10-loop/Size}{10}
\pgfkeyssetvalue{/bcc/Execution-Outcomes/well-formed-3-name-error/Size}{3}
\pgfkeyssetvalue{/bcc/Execution-Outcomes/well-formed-3-type-error/Size}{3}
\pgfkeyssetvalue{/bcc/Execution-Outcomes/well-formed-3-success/Size}{3}
\pgfkeyssetvalue{/bcc/Execution-Outcomes/well-formed-5-name-error/Size}{5}
\pgfkeyssetvalue{/bcc/Execution-Outcomes/well-formed-5-type-error/Size}{5}
\pgfkeyssetvalue{/bcc/Execution-Outcomes/well-formed-5-failure/Size}{5}
\pgfkeyssetvalue{/bcc/Execution-Outcomes/well-formed-5-success/Size}{5}
\pgfkeyssetvalue{/bcc/Execution-Outcomes/well-formed-7-name-error/Size}{7}
\pgfkeyssetvalue{/bcc/Execution-Outcomes/well-formed-7-success/Size}{7}
\pgfkeyssetvalue{/bcc/Execution-Outcomes/well-formed-7-type-error/Size}{7}
\pgfkeyssetvalue{/bcc/Execution-Outcomes/well-formed-7-failure/Size}{7}
\pgfkeyssetvalue{/bcc/Execution-Outcomes/well-formed-7-loop/Size}{7}
\pgfkeyssetvalue{/bcc/Execution-Outcomes/well-named-10-type-error/Size}{10}
\pgfkeyssetvalue{/bcc/Execution-Outcomes/well-named-10-success/Size}{10}
\pgfkeyssetvalue{/bcc/Execution-Outcomes/well-named-10-failure/Size}{10}
\pgfkeyssetvalue{/bcc/Execution-Outcomes/well-named-10-loop/Size}{10}
\pgfkeyssetvalue{/bcc/Execution-Outcomes/well-named-3-type-error/Size}{3}
\pgfkeyssetvalue{/bcc/Execution-Outcomes/well-named-3-failure/Size}{3}
\pgfkeyssetvalue{/bcc/Execution-Outcomes/well-named-3-loop/Size}{3}
\pgfkeyssetvalue{/bcc/Execution-Outcomes/well-named-3-success/Size}{3}
\pgfkeyssetvalue{/bcc/Execution-Outcomes/well-named-5-type-error/Size}{5}
\pgfkeyssetvalue{/bcc/Execution-Outcomes/well-named-5-success/Size}{5}
\pgfkeyssetvalue{/bcc/Execution-Outcomes/well-named-5-failure/Size}{5}
\pgfkeyssetvalue{/bcc/Execution-Outcomes/well-named-5-loop/Size}{5}
\pgfkeyssetvalue{/bcc/Execution-Outcomes/well-named-7-type-error/Size}{7}
\pgfkeyssetvalue{/bcc/Execution-Outcomes/well-named-7-success/Size}{7}
\pgfkeyssetvalue{/bcc/Execution-Outcomes/well-named-7-failure/Size}{7}
\pgfkeyssetvalue{/bcc/Execution-Outcomes/well-named-7-loop/Size}{7}
\pgfkeyssetvalue{/bcc/Execution-Outcomes/well-typed-10-failure/Size}{10}
\pgfkeyssetvalue{/bcc/Execution-Outcomes/well-typed-10-loop/Size}{10}
\pgfkeyssetvalue{/bcc/Execution-Outcomes/well-typed-10-timeout/Size}{10}
\pgfkeyssetvalue{/bcc/Execution-Outcomes/well-typed-10-success/Size}{10}
\pgfkeyssetvalue{/bcc/Execution-Outcomes/well-typed-3-failure/Size}{3}
\pgfkeyssetvalue{/bcc/Execution-Outcomes/well-typed-3-loop/Size}{3}
\pgfkeyssetvalue{/bcc/Execution-Outcomes/well-typed-3-success/Size}{3}
\pgfkeyssetvalue{/bcc/Execution-Outcomes/well-typed-3-timeout/Size}{3}
\pgfkeyssetvalue{/bcc/Execution-Outcomes/well-typed-5-failure/Size}{5}
\pgfkeyssetvalue{/bcc/Execution-Outcomes/well-typed-5-loop/Size}{5}
\pgfkeyssetvalue{/bcc/Execution-Outcomes/well-typed-5-success/Size}{5}
\pgfkeyssetvalue{/bcc/Execution-Outcomes/well-typed-5-timeout/Size}{5}
\pgfkeyssetvalue{/bcc/Execution-Outcomes/well-typed-7-loop/Size}{7}
\pgfkeyssetvalue{/bcc/Execution-Outcomes/well-typed-7-failure/Size}{7}
\pgfkeyssetvalue{/bcc/Execution-Outcomes/well-typed-7-success/Size}{7}
\pgfkeyssetvalue{/bcc/Execution-Outcomes/well-typed-7-timeout/Size}{7}
\pgfkeyssetvalue{/bcc/Execution-Outcomes/well-formed-10-name-error/Execution-Outcome}{name-error}
\pgfkeyssetvalue{/bcc/Execution-Outcomes/well-formed-10-type-error/Execution-Outcome}{type-error}
\pgfkeyssetvalue{/bcc/Execution-Outcomes/well-formed-10-success/Execution-Outcome}{success}
\pgfkeyssetvalue{/bcc/Execution-Outcomes/well-formed-10-failure/Execution-Outcome}{failure}
\pgfkeyssetvalue{/bcc/Execution-Outcomes/well-formed-10-loop/Execution-Outcome}{loop}
\pgfkeyssetvalue{/bcc/Execution-Outcomes/well-formed-3-name-error/Execution-Outcome}{name-error}
\pgfkeyssetvalue{/bcc/Execution-Outcomes/well-formed-3-type-error/Execution-Outcome}{type-error}
\pgfkeyssetvalue{/bcc/Execution-Outcomes/well-formed-3-success/Execution-Outcome}{success}
\pgfkeyssetvalue{/bcc/Execution-Outcomes/well-formed-5-name-error/Execution-Outcome}{name-error}
\pgfkeyssetvalue{/bcc/Execution-Outcomes/well-formed-5-type-error/Execution-Outcome}{type-error}
\pgfkeyssetvalue{/bcc/Execution-Outcomes/well-formed-5-failure/Execution-Outcome}{failure}
\pgfkeyssetvalue{/bcc/Execution-Outcomes/well-formed-5-success/Execution-Outcome}{success}
\pgfkeyssetvalue{/bcc/Execution-Outcomes/well-formed-7-name-error/Execution-Outcome}{name-error}
\pgfkeyssetvalue{/bcc/Execution-Outcomes/well-formed-7-success/Execution-Outcome}{success}
\pgfkeyssetvalue{/bcc/Execution-Outcomes/well-formed-7-type-error/Execution-Outcome}{type-error}
\pgfkeyssetvalue{/bcc/Execution-Outcomes/well-formed-7-failure/Execution-Outcome}{failure}
\pgfkeyssetvalue{/bcc/Execution-Outcomes/well-formed-7-loop/Execution-Outcome}{loop}
\pgfkeyssetvalue{/bcc/Execution-Outcomes/well-named-10-type-error/Execution-Outcome}{type-error}
\pgfkeyssetvalue{/bcc/Execution-Outcomes/well-named-10-success/Execution-Outcome}{success}
\pgfkeyssetvalue{/bcc/Execution-Outcomes/well-named-10-failure/Execution-Outcome}{failure}
\pgfkeyssetvalue{/bcc/Execution-Outcomes/well-named-10-loop/Execution-Outcome}{loop}
\pgfkeyssetvalue{/bcc/Execution-Outcomes/well-named-3-type-error/Execution-Outcome}{type-error}
\pgfkeyssetvalue{/bcc/Execution-Outcomes/well-named-3-failure/Execution-Outcome}{failure}
\pgfkeyssetvalue{/bcc/Execution-Outcomes/well-named-3-loop/Execution-Outcome}{loop}
\pgfkeyssetvalue{/bcc/Execution-Outcomes/well-named-3-success/Execution-Outcome}{success}
\pgfkeyssetvalue{/bcc/Execution-Outcomes/well-named-5-type-error/Execution-Outcome}{type-error}
\pgfkeyssetvalue{/bcc/Execution-Outcomes/well-named-5-success/Execution-Outcome}{success}
\pgfkeyssetvalue{/bcc/Execution-Outcomes/well-named-5-failure/Execution-Outcome}{failure}
\pgfkeyssetvalue{/bcc/Execution-Outcomes/well-named-5-loop/Execution-Outcome}{loop}
\pgfkeyssetvalue{/bcc/Execution-Outcomes/well-named-7-type-error/Execution-Outcome}{type-error}
\pgfkeyssetvalue{/bcc/Execution-Outcomes/well-named-7-success/Execution-Outcome}{success}
\pgfkeyssetvalue{/bcc/Execution-Outcomes/well-named-7-failure/Execution-Outcome}{failure}
\pgfkeyssetvalue{/bcc/Execution-Outcomes/well-named-7-loop/Execution-Outcome}{loop}
\pgfkeyssetvalue{/bcc/Execution-Outcomes/well-typed-10-failure/Execution-Outcome}{failure}
\pgfkeyssetvalue{/bcc/Execution-Outcomes/well-typed-10-loop/Execution-Outcome}{loop}
\pgfkeyssetvalue{/bcc/Execution-Outcomes/well-typed-10-timeout/Execution-Outcome}{timeout}
\pgfkeyssetvalue{/bcc/Execution-Outcomes/well-typed-10-success/Execution-Outcome}{success}
\pgfkeyssetvalue{/bcc/Execution-Outcomes/well-typed-3-failure/Execution-Outcome}{failure}
\pgfkeyssetvalue{/bcc/Execution-Outcomes/well-typed-3-loop/Execution-Outcome}{loop}
\pgfkeyssetvalue{/bcc/Execution-Outcomes/well-typed-3-success/Execution-Outcome}{success}
\pgfkeyssetvalue{/bcc/Execution-Outcomes/well-typed-3-timeout/Execution-Outcome}{timeout}
\pgfkeyssetvalue{/bcc/Execution-Outcomes/well-typed-5-failure/Execution-Outcome}{failure}
\pgfkeyssetvalue{/bcc/Execution-Outcomes/well-typed-5-loop/Execution-Outcome}{loop}
\pgfkeyssetvalue{/bcc/Execution-Outcomes/well-typed-5-success/Execution-Outcome}{success}
\pgfkeyssetvalue{/bcc/Execution-Outcomes/well-typed-5-timeout/Execution-Outcome}{timeout}
\pgfkeyssetvalue{/bcc/Execution-Outcomes/well-typed-7-loop/Execution-Outcome}{loop}
\pgfkeyssetvalue{/bcc/Execution-Outcomes/well-typed-7-failure/Execution-Outcome}{failure}
\pgfkeyssetvalue{/bcc/Execution-Outcomes/well-typed-7-success/Execution-Outcome}{success}
\pgfkeyssetvalue{/bcc/Execution-Outcomes/well-typed-7-timeout/Execution-Outcome}{timeout}
\pgfkeyssetvalue{/bcc/Execution-Outcomes/well-formed-10-name-error/Count}{499838}
\pgfkeyssetvalue{/bcc/Execution-Outcomes/well-formed-10-type-error/Count}{142}
\pgfkeyssetvalue{/bcc/Execution-Outcomes/well-formed-10-success/Count}{14}
\pgfkeyssetvalue{/bcc/Execution-Outcomes/well-formed-10-failure/Count}{5}
\pgfkeyssetvalue{/bcc/Execution-Outcomes/well-formed-10-loop/Count}{1}
\pgfkeyssetvalue{/bcc/Execution-Outcomes/well-formed-3-name-error/Count}{99617}
\pgfkeyssetvalue{/bcc/Execution-Outcomes/well-formed-3-type-error/Count}{381}
\pgfkeyssetvalue{/bcc/Execution-Outcomes/well-formed-3-success/Count}{2}
\pgfkeyssetvalue{/bcc/Execution-Outcomes/well-formed-5-name-error/Count}{199917}
\pgfkeyssetvalue{/bcc/Execution-Outcomes/well-formed-5-type-error/Count}{79}
\pgfkeyssetvalue{/bcc/Execution-Outcomes/well-formed-5-failure/Count}{2}
\pgfkeyssetvalue{/bcc/Execution-Outcomes/well-formed-5-success/Count}{2}
\pgfkeyssetvalue{/bcc/Execution-Outcomes/well-formed-7-name-error/Count}{199929}
\pgfkeyssetvalue{/bcc/Execution-Outcomes/well-formed-7-success/Count}{6}
\pgfkeyssetvalue{/bcc/Execution-Outcomes/well-formed-7-type-error/Count}{61}
\pgfkeyssetvalue{/bcc/Execution-Outcomes/well-formed-7-failure/Count}{3}
\pgfkeyssetvalue{/bcc/Execution-Outcomes/well-formed-7-loop/Count}{1}
\pgfkeyssetvalue{/bcc/Execution-Outcomes/well-named-10-type-error/Count}{499902}
\pgfkeyssetvalue{/bcc/Execution-Outcomes/well-named-10-success/Count}{73}
\pgfkeyssetvalue{/bcc/Execution-Outcomes/well-named-10-failure/Count}{18}
\pgfkeyssetvalue{/bcc/Execution-Outcomes/well-named-10-loop/Count}{7}
\pgfkeyssetvalue{/bcc/Execution-Outcomes/well-named-3-type-error/Count}{99938}
\pgfkeyssetvalue{/bcc/Execution-Outcomes/well-named-3-failure/Count}{25}
\pgfkeyssetvalue{/bcc/Execution-Outcomes/well-named-3-loop/Count}{14}
\pgfkeyssetvalue{/bcc/Execution-Outcomes/well-named-3-success/Count}{23}
\pgfkeyssetvalue{/bcc/Execution-Outcomes/well-named-5-type-error/Count}{199969}
\pgfkeyssetvalue{/bcc/Execution-Outcomes/well-named-5-success/Count}{21}
\pgfkeyssetvalue{/bcc/Execution-Outcomes/well-named-5-failure/Count}{9}
\pgfkeyssetvalue{/bcc/Execution-Outcomes/well-named-5-loop/Count}{1}
\pgfkeyssetvalue{/bcc/Execution-Outcomes/well-named-7-type-error/Count}{199965}
\pgfkeyssetvalue{/bcc/Execution-Outcomes/well-named-7-success/Count}{27}
\pgfkeyssetvalue{/bcc/Execution-Outcomes/well-named-7-failure/Count}{6}
\pgfkeyssetvalue{/bcc/Execution-Outcomes/well-named-7-loop/Count}{2}
\pgfkeyssetvalue{/bcc/Execution-Outcomes/well-typed-10-failure/Count}{341674}
\pgfkeyssetvalue{/bcc/Execution-Outcomes/well-typed-10-loop/Count}{155942}
\pgfkeyssetvalue{/bcc/Execution-Outcomes/well-typed-10-timeout/Count}{1074}
\pgfkeyssetvalue{/bcc/Execution-Outcomes/well-typed-10-success/Count}{1310}
\pgfkeyssetvalue{/bcc/Execution-Outcomes/well-typed-3-failure/Count}{64888}
\pgfkeyssetvalue{/bcc/Execution-Outcomes/well-typed-3-loop/Count}{31446}
\pgfkeyssetvalue{/bcc/Execution-Outcomes/well-typed-3-success/Count}{3664}
\pgfkeyssetvalue{/bcc/Execution-Outcomes/well-typed-3-timeout/Count}{2}
\pgfkeyssetvalue{/bcc/Execution-Outcomes/well-typed-5-failure/Count}{130578}
\pgfkeyssetvalue{/bcc/Execution-Outcomes/well-typed-5-loop/Count}{67661}
\pgfkeyssetvalue{/bcc/Execution-Outcomes/well-typed-5-success/Count}{1513}
\pgfkeyssetvalue{/bcc/Execution-Outcomes/well-typed-5-timeout/Count}{248}
\pgfkeyssetvalue{/bcc/Execution-Outcomes/well-typed-7-loop/Count}{68746}
\pgfkeyssetvalue{/bcc/Execution-Outcomes/well-typed-7-failure/Count}{130200}
\pgfkeyssetvalue{/bcc/Execution-Outcomes/well-typed-7-success/Count}{675}
\pgfkeyssetvalue{/bcc/Execution-Outcomes/well-typed-7-timeout/Count}{379}

\pgfkeyssetvalue{/bcc/Verification-Outcomes/name-error/Verification-Outcome}{name-error}
\pgfkeyssetvalue{/bcc/Verification-Outcomes/type-error/Verification-Outcome}{type-error}
\pgfkeyssetvalue{/bcc/Verification-Outcomes/success/Verification-Outcome}{success}
\pgfkeyssetvalue{/bcc/Verification-Outcomes/failure/Verification-Outcome}{failure}
\pgfkeyssetvalue{/bcc/Verification-Outcomes/other/Verification-Outcome}{other}
\pgfkeyssetvalue{/bcc/Verification-Outcomes/name-error/Count}{999301}
\pgfkeyssetvalue{/bcc/Verification-Outcomes/type-error/Count}{1000437}
\pgfkeyssetvalue{/bcc/Verification-Outcomes/success/Count}{267012}
\pgfkeyssetvalue{/bcc/Verification-Outcomes/failure/Count}{733224}
\pgfkeyssetvalue{/bcc/Verification-Outcomes/other/Count}{26}
\pgfkeyssetvalue{/bcc/Verification-Outcomes/name-error/Percentage}{33.31003333333334}
\pgfkeyssetvalue{/bcc/Verification-Outcomes/type-error/Percentage}{33.3479}
\pgfkeyssetvalue{/bcc/Verification-Outcomes/success/Percentage}{8.9004}
\pgfkeyssetvalue{/bcc/Verification-Outcomes/failure/Percentage}{24.4408}
\pgfkeyssetvalue{/bcc/Verification-Outcomes/other/Percentage}{0.0008666666666666666}

\pgfkeyssetvalue{/bcc/Verification-Outcomes/well-formed-name-error/Group}{well-formed}
\pgfkeyssetvalue{/bcc/Verification-Outcomes/well-formed-type-error/Group}{well-formed}
\pgfkeyssetvalue{/bcc/Verification-Outcomes/well-formed-success/Group}{well-formed}
\pgfkeyssetvalue{/bcc/Verification-Outcomes/well-formed-failure/Group}{well-formed}
\pgfkeyssetvalue{/bcc/Verification-Outcomes/well-named-type-error/Group}{well-named}
\pgfkeyssetvalue{/bcc/Verification-Outcomes/well-named-success/Group}{well-named}
\pgfkeyssetvalue{/bcc/Verification-Outcomes/well-named-failure/Group}{well-named}
\pgfkeyssetvalue{/bcc/Verification-Outcomes/well-typed-failure/Group}{well-typed}
\pgfkeyssetvalue{/bcc/Verification-Outcomes/well-typed-success/Group}{well-typed}
\pgfkeyssetvalue{/bcc/Verification-Outcomes/well-typed-other/Group}{well-typed}
\pgfkeyssetvalue{/bcc/Verification-Outcomes/well-formed-name-error/Verification-Outcome}{name-error}
\pgfkeyssetvalue{/bcc/Verification-Outcomes/well-formed-type-error/Verification-Outcome}{type-error}
\pgfkeyssetvalue{/bcc/Verification-Outcomes/well-formed-success/Verification-Outcome}{success}
\pgfkeyssetvalue{/bcc/Verification-Outcomes/well-formed-failure/Verification-Outcome}{failure}
\pgfkeyssetvalue{/bcc/Verification-Outcomes/well-named-type-error/Verification-Outcome}{type-error}
\pgfkeyssetvalue{/bcc/Verification-Outcomes/well-named-success/Verification-Outcome}{success}
\pgfkeyssetvalue{/bcc/Verification-Outcomes/well-named-failure/Verification-Outcome}{failure}
\pgfkeyssetvalue{/bcc/Verification-Outcomes/well-typed-failure/Verification-Outcome}{failure}
\pgfkeyssetvalue{/bcc/Verification-Outcomes/well-typed-success/Verification-Outcome}{success}
\pgfkeyssetvalue{/bcc/Verification-Outcomes/well-typed-other/Verification-Outcome}{other}
\pgfkeyssetvalue{/bcc/Verification-Outcomes/well-formed-name-error/Count}{999301}
\pgfkeyssetvalue{/bcc/Verification-Outcomes/well-formed-type-error/Count}{663}
\pgfkeyssetvalue{/bcc/Verification-Outcomes/well-formed-success/Count}{26}
\pgfkeyssetvalue{/bcc/Verification-Outcomes/well-formed-failure/Count}{10}
\pgfkeyssetvalue{/bcc/Verification-Outcomes/well-named-type-error/Count}{999774}
\pgfkeyssetvalue{/bcc/Verification-Outcomes/well-named-success/Count}{168}
\pgfkeyssetvalue{/bcc/Verification-Outcomes/well-named-failure/Count}{58}
\pgfkeyssetvalue{/bcc/Verification-Outcomes/well-typed-failure/Count}{733156}
\pgfkeyssetvalue{/bcc/Verification-Outcomes/well-typed-success/Count}{266818}
\pgfkeyssetvalue{/bcc/Verification-Outcomes/well-typed-other/Count}{26}
\pgfkeyssetvalue{/bcc/Verification-Outcomes/well-formed-name-error/Percentage}{99.9301}
\pgfkeyssetvalue{/bcc/Verification-Outcomes/well-formed-type-error/Percentage}{0.0663}
\pgfkeyssetvalue{/bcc/Verification-Outcomes/well-formed-success/Percentage}{0.0026}
\pgfkeyssetvalue{/bcc/Verification-Outcomes/well-formed-failure/Percentage}{0.001}
\pgfkeyssetvalue{/bcc/Verification-Outcomes/well-named-type-error/Percentage}{99.9774}
\pgfkeyssetvalue{/bcc/Verification-Outcomes/well-named-success/Percentage}{0.0168}
\pgfkeyssetvalue{/bcc/Verification-Outcomes/well-named-failure/Percentage}{0.0058}
\pgfkeyssetvalue{/bcc/Verification-Outcomes/well-typed-failure/Percentage}{73.3156}
\pgfkeyssetvalue{/bcc/Verification-Outcomes/well-typed-success/Percentage}{26.6818}
\pgfkeyssetvalue{/bcc/Verification-Outcomes/well-typed-other/Percentage}{0.0026}

\pgfkeyssetvalue{/bcc/Verification-Outcomes/well-formed-10-name-error/Group}{well-formed}
\pgfkeyssetvalue{/bcc/Verification-Outcomes/well-formed-10-type-error/Group}{well-formed}
\pgfkeyssetvalue{/bcc/Verification-Outcomes/well-formed-10-success/Group}{well-formed}
\pgfkeyssetvalue{/bcc/Verification-Outcomes/well-formed-10-failure/Group}{well-formed}
\pgfkeyssetvalue{/bcc/Verification-Outcomes/well-formed-3-name-error/Group}{well-formed}
\pgfkeyssetvalue{/bcc/Verification-Outcomes/well-formed-3-type-error/Group}{well-formed}
\pgfkeyssetvalue{/bcc/Verification-Outcomes/well-formed-3-success/Group}{well-formed}
\pgfkeyssetvalue{/bcc/Verification-Outcomes/well-formed-5-name-error/Group}{well-formed}
\pgfkeyssetvalue{/bcc/Verification-Outcomes/well-formed-5-type-error/Group}{well-formed}
\pgfkeyssetvalue{/bcc/Verification-Outcomes/well-formed-5-failure/Group}{well-formed}
\pgfkeyssetvalue{/bcc/Verification-Outcomes/well-formed-5-success/Group}{well-formed}
\pgfkeyssetvalue{/bcc/Verification-Outcomes/well-formed-7-name-error/Group}{well-formed}
\pgfkeyssetvalue{/bcc/Verification-Outcomes/well-formed-7-success/Group}{well-formed}
\pgfkeyssetvalue{/bcc/Verification-Outcomes/well-formed-7-type-error/Group}{well-formed}
\pgfkeyssetvalue{/bcc/Verification-Outcomes/well-formed-7-failure/Group}{well-formed}
\pgfkeyssetvalue{/bcc/Verification-Outcomes/well-named-10-type-error/Group}{well-named}
\pgfkeyssetvalue{/bcc/Verification-Outcomes/well-named-10-success/Group}{well-named}
\pgfkeyssetvalue{/bcc/Verification-Outcomes/well-named-10-failure/Group}{well-named}
\pgfkeyssetvalue{/bcc/Verification-Outcomes/well-named-3-type-error/Group}{well-named}
\pgfkeyssetvalue{/bcc/Verification-Outcomes/well-named-3-failure/Group}{well-named}
\pgfkeyssetvalue{/bcc/Verification-Outcomes/well-named-3-success/Group}{well-named}
\pgfkeyssetvalue{/bcc/Verification-Outcomes/well-named-5-type-error/Group}{well-named}
\pgfkeyssetvalue{/bcc/Verification-Outcomes/well-named-5-success/Group}{well-named}
\pgfkeyssetvalue{/bcc/Verification-Outcomes/well-named-5-failure/Group}{well-named}
\pgfkeyssetvalue{/bcc/Verification-Outcomes/well-named-7-type-error/Group}{well-named}
\pgfkeyssetvalue{/bcc/Verification-Outcomes/well-named-7-success/Group}{well-named}
\pgfkeyssetvalue{/bcc/Verification-Outcomes/well-named-7-failure/Group}{well-named}
\pgfkeyssetvalue{/bcc/Verification-Outcomes/well-typed-10-failure/Group}{well-typed}
\pgfkeyssetvalue{/bcc/Verification-Outcomes/well-typed-10-success/Group}{well-typed}
\pgfkeyssetvalue{/bcc/Verification-Outcomes/well-typed-10-other/Group}{well-typed}
\pgfkeyssetvalue{/bcc/Verification-Outcomes/well-typed-3-failure/Group}{well-typed}
\pgfkeyssetvalue{/bcc/Verification-Outcomes/well-typed-3-success/Group}{well-typed}
\pgfkeyssetvalue{/bcc/Verification-Outcomes/well-typed-5-failure/Group}{well-typed}
\pgfkeyssetvalue{/bcc/Verification-Outcomes/well-typed-5-success/Group}{well-typed}
\pgfkeyssetvalue{/bcc/Verification-Outcomes/well-typed-7-success/Group}{well-typed}
\pgfkeyssetvalue{/bcc/Verification-Outcomes/well-typed-7-failure/Group}{well-typed}
\pgfkeyssetvalue{/bcc/Verification-Outcomes/well-typed-7-other/Group}{well-typed}
\pgfkeyssetvalue{/bcc/Verification-Outcomes/well-formed-10-name-error/Size}{10}
\pgfkeyssetvalue{/bcc/Verification-Outcomes/well-formed-10-type-error/Size}{10}
\pgfkeyssetvalue{/bcc/Verification-Outcomes/well-formed-10-success/Size}{10}
\pgfkeyssetvalue{/bcc/Verification-Outcomes/well-formed-10-failure/Size}{10}
\pgfkeyssetvalue{/bcc/Verification-Outcomes/well-formed-3-name-error/Size}{3}
\pgfkeyssetvalue{/bcc/Verification-Outcomes/well-formed-3-type-error/Size}{3}
\pgfkeyssetvalue{/bcc/Verification-Outcomes/well-formed-3-success/Size}{3}
\pgfkeyssetvalue{/bcc/Verification-Outcomes/well-formed-5-name-error/Size}{5}
\pgfkeyssetvalue{/bcc/Verification-Outcomes/well-formed-5-type-error/Size}{5}
\pgfkeyssetvalue{/bcc/Verification-Outcomes/well-formed-5-failure/Size}{5}
\pgfkeyssetvalue{/bcc/Verification-Outcomes/well-formed-5-success/Size}{5}
\pgfkeyssetvalue{/bcc/Verification-Outcomes/well-formed-7-name-error/Size}{7}
\pgfkeyssetvalue{/bcc/Verification-Outcomes/well-formed-7-success/Size}{7}
\pgfkeyssetvalue{/bcc/Verification-Outcomes/well-formed-7-type-error/Size}{7}
\pgfkeyssetvalue{/bcc/Verification-Outcomes/well-formed-7-failure/Size}{7}
\pgfkeyssetvalue{/bcc/Verification-Outcomes/well-named-10-type-error/Size}{10}
\pgfkeyssetvalue{/bcc/Verification-Outcomes/well-named-10-success/Size}{10}
\pgfkeyssetvalue{/bcc/Verification-Outcomes/well-named-10-failure/Size}{10}
\pgfkeyssetvalue{/bcc/Verification-Outcomes/well-named-3-type-error/Size}{3}
\pgfkeyssetvalue{/bcc/Verification-Outcomes/well-named-3-failure/Size}{3}
\pgfkeyssetvalue{/bcc/Verification-Outcomes/well-named-3-success/Size}{3}
\pgfkeyssetvalue{/bcc/Verification-Outcomes/well-named-5-type-error/Size}{5}
\pgfkeyssetvalue{/bcc/Verification-Outcomes/well-named-5-success/Size}{5}
\pgfkeyssetvalue{/bcc/Verification-Outcomes/well-named-5-failure/Size}{5}
\pgfkeyssetvalue{/bcc/Verification-Outcomes/well-named-7-type-error/Size}{7}
\pgfkeyssetvalue{/bcc/Verification-Outcomes/well-named-7-success/Size}{7}
\pgfkeyssetvalue{/bcc/Verification-Outcomes/well-named-7-failure/Size}{7}
\pgfkeyssetvalue{/bcc/Verification-Outcomes/well-typed-10-failure/Size}{10}
\pgfkeyssetvalue{/bcc/Verification-Outcomes/well-typed-10-success/Size}{10}
\pgfkeyssetvalue{/bcc/Verification-Outcomes/well-typed-10-other/Size}{10}
\pgfkeyssetvalue{/bcc/Verification-Outcomes/well-typed-3-failure/Size}{3}
\pgfkeyssetvalue{/bcc/Verification-Outcomes/well-typed-3-success/Size}{3}
\pgfkeyssetvalue{/bcc/Verification-Outcomes/well-typed-5-failure/Size}{5}
\pgfkeyssetvalue{/bcc/Verification-Outcomes/well-typed-5-success/Size}{5}
\pgfkeyssetvalue{/bcc/Verification-Outcomes/well-typed-7-success/Size}{7}
\pgfkeyssetvalue{/bcc/Verification-Outcomes/well-typed-7-failure/Size}{7}
\pgfkeyssetvalue{/bcc/Verification-Outcomes/well-typed-7-other/Size}{7}
\pgfkeyssetvalue{/bcc/Verification-Outcomes/well-formed-10-name-error/Verification-Outcome}{name-error}
\pgfkeyssetvalue{/bcc/Verification-Outcomes/well-formed-10-type-error/Verification-Outcome}{type-error}
\pgfkeyssetvalue{/bcc/Verification-Outcomes/well-formed-10-success/Verification-Outcome}{success}
\pgfkeyssetvalue{/bcc/Verification-Outcomes/well-formed-10-failure/Verification-Outcome}{failure}
\pgfkeyssetvalue{/bcc/Verification-Outcomes/well-formed-3-name-error/Verification-Outcome}{name-error}
\pgfkeyssetvalue{/bcc/Verification-Outcomes/well-formed-3-type-error/Verification-Outcome}{type-error}
\pgfkeyssetvalue{/bcc/Verification-Outcomes/well-formed-3-success/Verification-Outcome}{success}
\pgfkeyssetvalue{/bcc/Verification-Outcomes/well-formed-5-name-error/Verification-Outcome}{name-error}
\pgfkeyssetvalue{/bcc/Verification-Outcomes/well-formed-5-type-error/Verification-Outcome}{type-error}
\pgfkeyssetvalue{/bcc/Verification-Outcomes/well-formed-5-failure/Verification-Outcome}{failure}
\pgfkeyssetvalue{/bcc/Verification-Outcomes/well-formed-5-success/Verification-Outcome}{success}
\pgfkeyssetvalue{/bcc/Verification-Outcomes/well-formed-7-name-error/Verification-Outcome}{name-error}
\pgfkeyssetvalue{/bcc/Verification-Outcomes/well-formed-7-success/Verification-Outcome}{success}
\pgfkeyssetvalue{/bcc/Verification-Outcomes/well-formed-7-type-error/Verification-Outcome}{type-error}
\pgfkeyssetvalue{/bcc/Verification-Outcomes/well-formed-7-failure/Verification-Outcome}{failure}
\pgfkeyssetvalue{/bcc/Verification-Outcomes/well-named-10-type-error/Verification-Outcome}{type-error}
\pgfkeyssetvalue{/bcc/Verification-Outcomes/well-named-10-success/Verification-Outcome}{success}
\pgfkeyssetvalue{/bcc/Verification-Outcomes/well-named-10-failure/Verification-Outcome}{failure}
\pgfkeyssetvalue{/bcc/Verification-Outcomes/well-named-3-type-error/Verification-Outcome}{type-error}
\pgfkeyssetvalue{/bcc/Verification-Outcomes/well-named-3-failure/Verification-Outcome}{failure}
\pgfkeyssetvalue{/bcc/Verification-Outcomes/well-named-3-success/Verification-Outcome}{success}
\pgfkeyssetvalue{/bcc/Verification-Outcomes/well-named-5-type-error/Verification-Outcome}{type-error}
\pgfkeyssetvalue{/bcc/Verification-Outcomes/well-named-5-success/Verification-Outcome}{success}
\pgfkeyssetvalue{/bcc/Verification-Outcomes/well-named-5-failure/Verification-Outcome}{failure}
\pgfkeyssetvalue{/bcc/Verification-Outcomes/well-named-7-type-error/Verification-Outcome}{type-error}
\pgfkeyssetvalue{/bcc/Verification-Outcomes/well-named-7-success/Verification-Outcome}{success}
\pgfkeyssetvalue{/bcc/Verification-Outcomes/well-named-7-failure/Verification-Outcome}{failure}
\pgfkeyssetvalue{/bcc/Verification-Outcomes/well-typed-10-failure/Verification-Outcome}{failure}
\pgfkeyssetvalue{/bcc/Verification-Outcomes/well-typed-10-success/Verification-Outcome}{success}
\pgfkeyssetvalue{/bcc/Verification-Outcomes/well-typed-10-other/Verification-Outcome}{other}
\pgfkeyssetvalue{/bcc/Verification-Outcomes/well-typed-3-failure/Verification-Outcome}{failure}
\pgfkeyssetvalue{/bcc/Verification-Outcomes/well-typed-3-success/Verification-Outcome}{success}
\pgfkeyssetvalue{/bcc/Verification-Outcomes/well-typed-5-failure/Verification-Outcome}{failure}
\pgfkeyssetvalue{/bcc/Verification-Outcomes/well-typed-5-success/Verification-Outcome}{success}
\pgfkeyssetvalue{/bcc/Verification-Outcomes/well-typed-7-success/Verification-Outcome}{success}
\pgfkeyssetvalue{/bcc/Verification-Outcomes/well-typed-7-failure/Verification-Outcome}{failure}
\pgfkeyssetvalue{/bcc/Verification-Outcomes/well-typed-7-other/Verification-Outcome}{other}
\pgfkeyssetvalue{/bcc/Verification-Outcomes/well-formed-10-name-error/Count}{499838}
\pgfkeyssetvalue{/bcc/Verification-Outcomes/well-formed-10-type-error/Count}{142}
\pgfkeyssetvalue{/bcc/Verification-Outcomes/well-formed-10-success/Count}{15}
\pgfkeyssetvalue{/bcc/Verification-Outcomes/well-formed-10-failure/Count}{5}
\pgfkeyssetvalue{/bcc/Verification-Outcomes/well-formed-3-name-error/Count}{99617}
\pgfkeyssetvalue{/bcc/Verification-Outcomes/well-formed-3-type-error/Count}{381}
\pgfkeyssetvalue{/bcc/Verification-Outcomes/well-formed-3-success/Count}{2}
\pgfkeyssetvalue{/bcc/Verification-Outcomes/well-formed-5-name-error/Count}{199917}
\pgfkeyssetvalue{/bcc/Verification-Outcomes/well-formed-5-type-error/Count}{79}
\pgfkeyssetvalue{/bcc/Verification-Outcomes/well-formed-5-failure/Count}{2}
\pgfkeyssetvalue{/bcc/Verification-Outcomes/well-formed-5-success/Count}{2}
\pgfkeyssetvalue{/bcc/Verification-Outcomes/well-formed-7-name-error/Count}{199929}
\pgfkeyssetvalue{/bcc/Verification-Outcomes/well-formed-7-success/Count}{7}
\pgfkeyssetvalue{/bcc/Verification-Outcomes/well-formed-7-type-error/Count}{61}
\pgfkeyssetvalue{/bcc/Verification-Outcomes/well-formed-7-failure/Count}{3}
\pgfkeyssetvalue{/bcc/Verification-Outcomes/well-named-10-type-error/Count}{499902}
\pgfkeyssetvalue{/bcc/Verification-Outcomes/well-named-10-success/Count}{80}
\pgfkeyssetvalue{/bcc/Verification-Outcomes/well-named-10-failure/Count}{18}
\pgfkeyssetvalue{/bcc/Verification-Outcomes/well-named-3-type-error/Count}{99938}
\pgfkeyssetvalue{/bcc/Verification-Outcomes/well-named-3-failure/Count}{25}
\pgfkeyssetvalue{/bcc/Verification-Outcomes/well-named-3-success/Count}{37}
\pgfkeyssetvalue{/bcc/Verification-Outcomes/well-named-5-type-error/Count}{199969}
\pgfkeyssetvalue{/bcc/Verification-Outcomes/well-named-5-success/Count}{22}
\pgfkeyssetvalue{/bcc/Verification-Outcomes/well-named-5-failure/Count}{9}
\pgfkeyssetvalue{/bcc/Verification-Outcomes/well-named-7-type-error/Count}{199965}
\pgfkeyssetvalue{/bcc/Verification-Outcomes/well-named-7-success/Count}{29}
\pgfkeyssetvalue{/bcc/Verification-Outcomes/well-named-7-failure/Count}{6}
\pgfkeyssetvalue{/bcc/Verification-Outcomes/well-typed-10-failure/Count}{386264}
\pgfkeyssetvalue{/bcc/Verification-Outcomes/well-typed-10-success/Count}{113711}
\pgfkeyssetvalue{/bcc/Verification-Outcomes/well-typed-10-other/Count}{25}
\pgfkeyssetvalue{/bcc/Verification-Outcomes/well-typed-3-failure/Count}{66040}
\pgfkeyssetvalue{/bcc/Verification-Outcomes/well-typed-3-success/Count}{33960}
\pgfkeyssetvalue{/bcc/Verification-Outcomes/well-typed-5-failure/Count}{137874}
\pgfkeyssetvalue{/bcc/Verification-Outcomes/well-typed-5-success/Count}{62126}
\pgfkeyssetvalue{/bcc/Verification-Outcomes/well-typed-7-success/Count}{57021}
\pgfkeyssetvalue{/bcc/Verification-Outcomes/well-typed-7-failure/Count}{142978}
\pgfkeyssetvalue{/bcc/Verification-Outcomes/well-typed-7-other/Count}{1}

\pgfkeyssetvalue{/bcc/Execution-vs-Verification/name-error-name-error/Execution-Outcome}{name-error}
\pgfkeyssetvalue{/bcc/Execution-vs-Verification/type-error-type-error/Execution-Outcome}{type-error}
\pgfkeyssetvalue{/bcc/Execution-vs-Verification/success-success/Execution-Outcome}{success}
\pgfkeyssetvalue{/bcc/Execution-vs-Verification/success-failure/Execution-Outcome}{success}
\pgfkeyssetvalue{/bcc/Execution-vs-Verification/failure-failure/Execution-Outcome}{failure}
\pgfkeyssetvalue{/bcc/Execution-vs-Verification/failure-other/Execution-Outcome}{failure}
\pgfkeyssetvalue{/bcc/Execution-vs-Verification/loop-success/Execution-Outcome}{loop}
\pgfkeyssetvalue{/bcc/Execution-vs-Verification/loop-failure/Execution-Outcome}{loop}
\pgfkeyssetvalue{/bcc/Execution-vs-Verification/loop-other/Execution-Outcome}{loop}
\pgfkeyssetvalue{/bcc/Execution-vs-Verification/timeout-success/Execution-Outcome}{timeout}
\pgfkeyssetvalue{/bcc/Execution-vs-Verification/timeout-failure/Execution-Outcome}{timeout}
\pgfkeyssetvalue{/bcc/Execution-vs-Verification/name-error-name-error/Verification-Outcome}{name-error}
\pgfkeyssetvalue{/bcc/Execution-vs-Verification/type-error-type-error/Verification-Outcome}{type-error}
\pgfkeyssetvalue{/bcc/Execution-vs-Verification/success-success/Verification-Outcome}{success}
\pgfkeyssetvalue{/bcc/Execution-vs-Verification/success-failure/Verification-Outcome}{failure}
\pgfkeyssetvalue{/bcc/Execution-vs-Verification/failure-failure/Verification-Outcome}{failure}
\pgfkeyssetvalue{/bcc/Execution-vs-Verification/failure-other/Verification-Outcome}{other}
\pgfkeyssetvalue{/bcc/Execution-vs-Verification/loop-success/Verification-Outcome}{success}
\pgfkeyssetvalue{/bcc/Execution-vs-Verification/loop-failure/Verification-Outcome}{failure}
\pgfkeyssetvalue{/bcc/Execution-vs-Verification/loop-other/Verification-Outcome}{other}
\pgfkeyssetvalue{/bcc/Execution-vs-Verification/timeout-success/Verification-Outcome}{success}
\pgfkeyssetvalue{/bcc/Execution-vs-Verification/timeout-failure/Verification-Outcome}{failure}
\pgfkeyssetvalue{/bcc/Execution-vs-Verification/name-error-name-error/Count}{999301}
\pgfkeyssetvalue{/bcc/Execution-vs-Verification/type-error-type-error/Count}{1000437}
\pgfkeyssetvalue{/bcc/Execution-vs-Verification/success-success/Count}{6981}
\pgfkeyssetvalue{/bcc/Execution-vs-Verification/success-failure/Count}{349}
\pgfkeyssetvalue{/bcc/Execution-vs-Verification/failure-failure/Count}{667397}
\pgfkeyssetvalue{/bcc/Execution-vs-Verification/failure-other/Count}{11}
\pgfkeyssetvalue{/bcc/Execution-vs-Verification/loop-success/Count}{258808}
\pgfkeyssetvalue{/bcc/Execution-vs-Verification/loop-failure/Count}{64998}
\pgfkeyssetvalue{/bcc/Execution-vs-Verification/loop-other/Count}{15}
\pgfkeyssetvalue{/bcc/Execution-vs-Verification/timeout-success/Count}{1223}
\pgfkeyssetvalue{/bcc/Execution-vs-Verification/timeout-failure/Count}{480}
\pgfkeyssetvalue{/bcc/Execution-vs-Verification/name-error-name-error/Percentage}{33.31003333333334}
\pgfkeyssetvalue{/bcc/Execution-vs-Verification/type-error-type-error/Percentage}{33.3479}
\pgfkeyssetvalue{/bcc/Execution-vs-Verification/success-success/Percentage}{0.23270000000000002}
\pgfkeyssetvalue{/bcc/Execution-vs-Verification/success-failure/Percentage}{0.011633333333333334}
\pgfkeyssetvalue{/bcc/Execution-vs-Verification/failure-failure/Percentage}{22.246566666666666}
\pgfkeyssetvalue{/bcc/Execution-vs-Verification/failure-other/Percentage}{0.00036666666666666667}
\pgfkeyssetvalue{/bcc/Execution-vs-Verification/loop-success/Percentage}{8.626933333333334}
\pgfkeyssetvalue{/bcc/Execution-vs-Verification/loop-failure/Percentage}{2.1666000000000003}
\pgfkeyssetvalue{/bcc/Execution-vs-Verification/loop-other/Percentage}{0.0005}
\pgfkeyssetvalue{/bcc/Execution-vs-Verification/timeout-success/Percentage}{0.040766666666666666}
\pgfkeyssetvalue{/bcc/Execution-vs-Verification/timeout-failure/Percentage}{0.016}

\pgfkeyssetvalue{/bcc/Execution-vs-Verification-Infer/loop-success/Execution-Outcome}{loop}
\pgfkeyssetvalue{/bcc/Execution-vs-Verification-Infer/loop-failure/Execution-Outcome}{loop}
\pgfkeyssetvalue{/bcc/Execution-vs-Verification-Infer/success-success/Execution-Outcome}{success}
\pgfkeyssetvalue{/bcc/Execution-vs-Verification-Infer/success-failure/Execution-Outcome}{success}
\pgfkeyssetvalue{/bcc/Execution-vs-Verification-Infer/loop-other/Execution-Outcome}{loop}
\pgfkeyssetvalue{/bcc/Execution-vs-Verification-Infer/name-error-name-error/Execution-Outcome}{name-error}
\pgfkeyssetvalue{/bcc/Execution-vs-Verification-Infer/type-error-type-error/Execution-Outcome}{type-error}
\pgfkeyssetvalue{/bcc/Execution-vs-Verification-Infer/failure-failure/Execution-Outcome}{failure}
\pgfkeyssetvalue{/bcc/Execution-vs-Verification-Infer/timeout-success/Execution-Outcome}{timeout}
\pgfkeyssetvalue{/bcc/Execution-vs-Verification-Infer/timeout-failure/Execution-Outcome}{timeout}
\pgfkeyssetvalue{/bcc/Execution-vs-Verification-Infer/failure-other/Execution-Outcome}{failure}
\pgfkeyssetvalue{/bcc/Execution-vs-Verification-Infer/loop-success/Verification-Outcome}{success}
\pgfkeyssetvalue{/bcc/Execution-vs-Verification-Infer/loop-failure/Verification-Outcome}{failure}
\pgfkeyssetvalue{/bcc/Execution-vs-Verification-Infer/success-success/Verification-Outcome}{success}
\pgfkeyssetvalue{/bcc/Execution-vs-Verification-Infer/success-failure/Verification-Outcome}{failure}
\pgfkeyssetvalue{/bcc/Execution-vs-Verification-Infer/loop-other/Verification-Outcome}{other}
\pgfkeyssetvalue{/bcc/Execution-vs-Verification-Infer/name-error-name-error/Verification-Outcome}{name-error}
\pgfkeyssetvalue{/bcc/Execution-vs-Verification-Infer/type-error-type-error/Verification-Outcome}{type-error}
\pgfkeyssetvalue{/bcc/Execution-vs-Verification-Infer/failure-failure/Verification-Outcome}{failure}
\pgfkeyssetvalue{/bcc/Execution-vs-Verification-Infer/timeout-success/Verification-Outcome}{success}
\pgfkeyssetvalue{/bcc/Execution-vs-Verification-Infer/timeout-failure/Verification-Outcome}{failure}
\pgfkeyssetvalue{/bcc/Execution-vs-Verification-Infer/failure-other/Verification-Outcome}{other}
\pgfkeyssetvalue{/bcc/Execution-vs-Verification-Infer/loop-success/Count}{309020}
\pgfkeyssetvalue{/bcc/Execution-vs-Verification-Infer/loop-failure/Count}{14785}
\pgfkeyssetvalue{/bcc/Execution-vs-Verification-Infer/success-success/Count}{7259}
\pgfkeyssetvalue{/bcc/Execution-vs-Verification-Infer/success-failure/Count}{71}
\pgfkeyssetvalue{/bcc/Execution-vs-Verification-Infer/loop-other/Count}{16}
\pgfkeyssetvalue{/bcc/Execution-vs-Verification-Infer/name-error-name-error/Count}{999301}
\pgfkeyssetvalue{/bcc/Execution-vs-Verification-Infer/type-error-type-error/Count}{1000437}
\pgfkeyssetvalue{/bcc/Execution-vs-Verification-Infer/failure-failure/Count}{667397}
\pgfkeyssetvalue{/bcc/Execution-vs-Verification-Infer/timeout-success/Count}{1223}
\pgfkeyssetvalue{/bcc/Execution-vs-Verification-Infer/timeout-failure/Count}{480}
\pgfkeyssetvalue{/bcc/Execution-vs-Verification-Infer/failure-other/Count}{11}
\pgfkeyssetvalue{/bcc/Execution-vs-Verification-Infer/loop-success/Percentage}{10.300666666666666}
\pgfkeyssetvalue{/bcc/Execution-vs-Verification-Infer/loop-failure/Percentage}{0.49283333333333335}
\pgfkeyssetvalue{/bcc/Execution-vs-Verification-Infer/success-success/Percentage}{0.24196666666666666}
\pgfkeyssetvalue{/bcc/Execution-vs-Verification-Infer/success-failure/Percentage}{0.0023666666666666667}
\pgfkeyssetvalue{/bcc/Execution-vs-Verification-Infer/loop-other/Percentage}{0.0005333333333333334}
\pgfkeyssetvalue{/bcc/Execution-vs-Verification-Infer/name-error-name-error/Percentage}{33.31003333333334}
\pgfkeyssetvalue{/bcc/Execution-vs-Verification-Infer/type-error-type-error/Percentage}{33.3479}
\pgfkeyssetvalue{/bcc/Execution-vs-Verification-Infer/failure-failure/Percentage}{22.246566666666666}
\pgfkeyssetvalue{/bcc/Execution-vs-Verification-Infer/timeout-success/Percentage}{0.040766666666666666}
\pgfkeyssetvalue{/bcc/Execution-vs-Verification-Infer/timeout-failure/Percentage}{0.016}
\pgfkeyssetvalue{/bcc/Execution-vs-Verification-Infer/failure-other/Percentage}{0.00036666666666666667}

\newcommand{\termrow}[2]{#2 & 
  \n{Term-Counts/#1-#2/Locals-Minimum}
  & \n[0]{Term-Counts/#1-#2/Locals-Median}
  & \n[0]{Term-Counts/#1-#2/Locals-Maximum}
  & \n[0]{Term-Counts/#1-#2/Stmt-Count-Minimum}
  & \n[0]{Term-Counts/#1-#2/Stmt-Count-Median}
  & \n[0]{Term-Counts/#1-#2/Stmt-Count-Maximum}
  & \n[0]{Term-Counts/#1-#2/ARI-Count-Minimum}
  & \n[0]{Term-Counts/#1-#2/ARI-Count-Median}
  & \n[0]{Term-Counts/#1-#2/ARI-Count-Maximum}
  & \n[0]{Term-Counts/#1-#2/LOG-Count-Minimum}
  & \n[0]{Term-Counts/#1-#2/LOG-Count-Median}
  & \n[0]{Term-Counts/#1-#2/LOG-Count-Maximum}
  & \n[0]{Term-Counts/#1-#2/CMP-Count-Minimum}
  & \n[0]{Term-Counts/#1-#2/CMP-Count-Median}
  & \n[0]{Term-Counts/#1-#2/CMP-Count-Maximum}
  & \n[0]{Term-Counts/#1-#2/LIT-Count-Minimum}
  & \n[0]{Term-Counts/#1-#2/LIT-Count-Median}
  & \n[0]{Term-Counts/#1-#2/LIT-Count-Maximum}
}

\newcommand{\execVerifRow}[2]{#2 & 
  \n{Execution-Outcomes/#1-#2-success/Count} &
  \n{Execution-Outcomes/#1-#2-failure/Count} &
  \n{Execution-Outcomes/#1-#2-loop/Count} &
  \n{Execution-Outcomes/#1-#2-timeout/Count} &
  \n{Execution-Outcomes/#1-#2-name-error/Count} &
  \n{Execution-Outcomes/#1-#2-type-error/Count} &
  \n{Verification-Outcomes/#1-#2-success/Count} &
  \n{Verification-Outcomes/#1-#2-failure/Count} &
  \n{Verification-Outcomes/#1-#2-other/Count} &  \n{Verification-Outcomes/#1-#2-name-error/Count} &
  \n{Verification-Outcomes/#1-#2-type-error/Count}
}

\newcommand{\execVerifGroupRow}[1]{\quad \#
  & \n{Execution-Outcomes/#1-success/Count}
  & \n{Execution-Outcomes/#1-failure/Count}
  & \n{Execution-Outcomes/#1-loop/Count}
  & \n{Execution-Outcomes/#1-timeout/Count}
  & \n{Execution-Outcomes/#1-name-error/Count}
  & \n{Execution-Outcomes/#1-type-error/Count}
  & \n{Verification-Outcomes/#1-success/Count}
  & \n{Verification-Outcomes/#1-failure/Count}
  & \n{Verification-Outcomes/#1-other/Count}  & \n{Verification-Outcomes/#1-name-error/Count}
  & \n{Verification-Outcomes/#1-type-error/Count}
  \\
  & \quad \%
  & \n[0]{Execution-Outcomes/#1-success/Percentage}
  & \n[0]{Execution-Outcomes/#1-failure/Percentage}
  & \n[0]{Execution-Outcomes/#1-loop/Percentage}
  & \n[0]{Execution-Outcomes/#1-timeout/Percentage}
  & \n[0]{Execution-Outcomes/#1-name-error/Percentage}
  & \n[0]{Execution-Outcomes/#1-type-error/Percentage}
  & \n[0]{Verification-Outcomes/#1-success/Percentage}
  & \n[0]{Verification-Outcomes/#1-failure/Percentage}
  & \n[0]{Verification-Outcomes/#1-other/Percentage}
  & \n[0]{Verification-Outcomes/#1-name-error/Percentage}
  & \n[0]{Verification-Outcomes/#1-type-error/Percentage}
  \\
}

\newcommand{\execVerifTotalRow}{\quad \#
  & \n{Execution-Outcomes/success/Count}
  & \n{Execution-Outcomes/failure/Count}
  & \n{Execution-Outcomes/loop/Count}
  & \n{Execution-Outcomes/timeout/Count}
  & \n{Execution-Outcomes/name-error/Count}
  & \n{Execution-Outcomes/type-error/Count}
  & \n{Verification-Outcomes/success/Count}
  & \n{Verification-Outcomes/failure/Count}
  & \n{Verification-Outcomes/other/Count}  & \n{Verification-Outcomes/name-error/Count}
  & \n{Verification-Outcomes/type-error/Count}
  \\
  & \quad \%
  & \n[0]{Execution-Outcomes/success/Percentage}
  & \n[0]{Execution-Outcomes/failure/Percentage}
  & \n[0]{Execution-Outcomes/loop/Percentage}
  & \n[0]{Execution-Outcomes/timeout/Percentage}
  & \n[0]{Execution-Outcomes/name-error/Percentage}
  & \n[0]{Execution-Outcomes/type-error/Percentage}
  & \n[0]{Verification-Outcomes/success/Percentage}
  & \n[0]{Verification-Outcomes/failure/Percentage}
  & \n[0]{Verification-Outcomes/other/Percentage}
  & \n[0]{Verification-Outcomes/name-error/Percentage}
  & \n[0]{Verification-Outcomes/type-error/Percentage}
  \\
}

\DeclareDocumentCommand{\bvpcell}{s m m}{\IfBooleanT{#1}{\cellcolor{nocolbg}}
  \n[0]{Execution-vs-Verification/#2-#3/Count}
  &
  \IfBooleanT{#1}{\cellcolor{nocolbg}}
  \n[2]{Execution-vs-Verification/#2-#3/Percentage}|
}

\usepackage{listings}

\usepackage{boogie}

\newcommand{\Bpl}[1]{\mbox{\lstinline[basicstyle=\ttfamily,language=boogie,breaklines=true]|#1|}}

\lstdefinelanguage{racket} {
  morekeywords=[1]{define, define-syntax, define-macro, lambda, define-stream, stream-lambda},
  morekeywords=[2]{
assert, while, main, true, false,
begin, call-with-current-continuation, call/cc,
    call-with-input-file, call-with-output-file, case, cond,
    do, else, for-each, if,
    let*, let, let-syntax, letrec, letrec-syntax,
    define-context, define-controller, Integer, Boolean, get, when-required, when-provided,
    maybe_publish, require, submod, or/c, ->, \#\%module-begin,
    always_publish, with-syntax, define-struct/contract, syntax-case,
    define/contract,
    let-values, let*-values,
    module, provide,
    and, or, not, delay, force,
    \#`, \#',
    \#lang, implement, begin-for-syntax, rename-out,
    quasiquote, quote, unquote, unquote-splicing,
    map, fold, syntax, syntax-rules, eval, environment, query },
  morekeywords=[3]{import, export},
  alsoletter={',`,-,/,>,<,\#,\%},
  morecomment=[l]{;},
moredelim=**[is][\color{light-gray}]{<<@<<}{>>@>>},
  moredelim=**[is][\itshape\color{OliveGreen}]{<<;<<}{>>;>>},
  morecomment=[s]{\#|}{|\#},
  sensitive=true,
}
\newcommand{\Rkt}[1]{\mbox{\lstinline[basicstyle=\ttfamily,language=racket,breaklines=true]|#1|}}

\usepackage{savesym}
\savesymbol{endnote}

\usepackage{enotez}

\restoresymbol{enotez}{endnote}

\setenotez{list-name={URL References}}
\setenotez{counter-format=alph}
\setenotez{mark-cs={\formatEndNoteMark}}
\newcommand{\formatEndNoteMark}[1]{\textsuperscript{[\color{blue}{\textsf{#1}}]}}

\DeclareInstance{enotez-list}{custom}{list}{list-type=itemize,heading=\section*{#1}}

\DeclareDocumentCommand{\urlcite}{s O{} m}
{\enotezendnote{\IfNoValueF{#2}{#2\xspace}\url{#3}}}

\let\llncssubparagraph\subparagraph
\let\subparagraph\paragraph
\usepackage[compact]{titlesec}
\let\subparagraph\llncssubparagraph

\newcommand{\bcc}{{\smaller[0.5]{\textsc{bcc}}}\xspace}
\newcommand{\bplsub}{{\smaller[0.5]{\mbox{\textsc{bpl}$_0$}}}\xspace}
\newcommand{\plt}{{\smaller[0.5]{\textsc{plt}~Red\-ex}}\xspace}

\ifdraft
\usepackage[colorinlistoftodos,textsize=scriptsize,obeyFinal,textwidth=33mm]{todonotes}

\DeclareDocumentCommand{\ReviewNote}{s o m O{white}}{\todo[color=#4,\IfBooleanTF{#1}{inline}{}]{\IfNoValueF{#2}{\textbf{#2:}\xspace}#3}
}
\else
\DeclareDocumentCommand{\ReviewNote}{s o m O{white}}{}
\fi

\DeclareDocumentCommand{\caf}{s m}{\IfBooleanTF{#1}{\ReviewNote*{#2}[yellow]}{\ReviewNote{#2}[yellow]}}
\DeclareDocumentCommand{\mpg}{s m}{\IfBooleanTF{#1}{\ReviewNote*{#2}[red!70!white]}{\ReviewNote{#2}[red!70!white]}}

\usepackage{wrapfig}

\begin{document}

\title{Model-Based Testing of an Intermediate Verifier\\ Using Executable Operational Semantics\ifblind\else\thanks{Work partially supported by SNF grant 200021-207919 (LastMile).}\fi}

\ifblind
  \author{Anonymous authors}
\else
\author{Lidia Losavio\inst{1}\thanks{These authors contributed equally; they are listed in alphabetical order.} \and Marco Paganoni\inst{1}\textsuperscript{$\ast\ast$} \and
Carlo A. Furia\inst{1}\orcidID{0000-0003-1040-3201}}
\authorrunning{L.\ Losavio et al.}

\institute{Software Institute, USI Università della Svizzera italiana, Lugano, Switzerland\\
\email{lidia.losavio@usi.ch}  $\qquad$ \email{marco.paganoni@usi.ch} $\qquad$ \url{bugcounting.net}}
\fi

\maketitle

\renewcommand{\leadsto}{\hookrightarrow}

\begin{abstract}
  Lightweight validation technique, such as those based on random testing,
  are sometimes practical alternatives to full formal verification---providing valuable benefits, such as finding bugs,
  without requiring a disproportionate effort.
  In fact, such validation techniques
  can be useful even for fully formally verified tools,
  by exercising the parts of a complex system that
  go beyond the reach of formal models.
  
  In this context, this paper introduces \bcc: a model-based testing
  technique for the Boogie intermediate verifier.
  \bcc combines the formalization of a small, deterministic
  subset of the Boogie language
  with the generative capabilities of the \plt language engineering framework.
  Basically, \bcc uses \plt to
  generate random Boogie programs, and to execute them according
  to a formal operational semantics;
  then, it runs the same programs through the Boogie verifier.
  Any inconsistency between the two executions
  (in \plt and with Boogie)
  may indicate a potential bug in Boogie's implementation.

  To understand whether \bcc can be useful in practice,
  we used it to generate three million Boogie programs.
  These experiments found \n[0]{perc/failures}[100]| of cases
  indicative of 
completeness failures (i.e., spurious verification failures)
  in Boogie's toolchain.
  These results indicate that lightweight analysis tools,
  such as those for model-based random testing,
  are also useful to test and validate formal verification tools such as Boogie.
\end{abstract}

\section{Introduction}
\label{sec:introduction}

Modern verification tools are complex pieces of software;
as such, they may contain defects of various kinds
that can affect their soundness, precision, performance, or usability.
Even if formally verifying a verifier's implementation is the ideal end goal,
it usually remains a daunting challenge that requires
massive amounts of expert effort~\cite{LinCTWR23,DardinierSPSM25}.
This motivates (also) developing \emph{lightweight} techniques~\cite{DBLP:conf/sat/BrummayerLB10,MansurCWZ20,DBLP:conf/kbse/BringolfW022,boogie-robustness-testing,DBLP:journals/pacmpl/Winterer024},
which do not provide absolute correctness guarantees
but can still find errors
or provide partial validation
with a high degree of automation.

In this vein, this paper describes
\bcc:\footnote{\bcc stands for ``\underline{B}oogie \underline{c}onsistency \underline{c}hecker''.}
a lightweight technique
to test the implementation of the Boogie intermediate verifier~\cite{Boogie}.
\bcc
relies on a formalization of \bplsub ---a small, deterministic subset of the Boogie language---encoded using the \plt framework~\cite{PLT-Redex}.
In a nutshell, the formal model
describes the syntax of \bplsub,
its typing rules,
and its execution semantics by means of symbolic reduction rules.
\bcc first uses the formal model in \plt to
generate random \bplsub programs.
Then, it executes the programs according to their operational semantics,
and also runs the Boogie verifier on them.
The outcome of Boogie's verification run
and of the execution in \plt should be \emph{consistent};
\bcc reports any inconsistency as a potential \emph{error}
  of Boogie .\footnote{
  In principle, an inconsistency may also be due to a bug
  in the operational semantics rules;
  in practice, \bplsub and its semantics
  are so much simpler than Boogie that
  it's overwhelmingly much more likely that it's
  Boogie's implementation that at fault.
  }

Since it is a best-effort validation technique,
\bcc cannot provide absolute guarantees of correctness.
\iflong
\else
  In particular,
\fi
Boogie's verification semantics and 
\bcc's operational semantics may be inconsistent in certain cases
without implications for correctness:
for example, executing a loop in \bcc's semantics
may time out, whereas Boogie analyzes loops symbolically
without checking whether they terminate.
Reconciling Boogie's verification semantics
and \bcc's executable operational semantics
is a key challenge addressed in this work---the first, to our knowledge, to apply \plt's semantic engineering
techniques to a verification language
(as opposed to a conventional programming language).

\iflong
\else O\fi ur
experiments demonstrate that \bcc can be practically useful
as a bug-finding and validation tool.
We generated three million syntactically different random Boogie programs
of different sizes and characteristics.
\bcc found \n{count/failures} cases
(\n[0]{perc/failures}[100]| of all generated programs)
of \emph{completeness} failures.
Even though \bcc only covers a small subset of the whole Boogie language (see \autoref{fig:bpl0-grammar}),
our approach shows that semantics engineering techniques,
such as those made available by \plt,
can be applied to test formal verification tools,
and help improve their reliability.

In summary, the paper makes the following main contributions:
\begin{enumerate*}
\item An executable operational semantics of \bplsub,
  a deterministic subset of the Boogie intermediate verification language,
  built using the \plt framework.
\item \bcc: a technique to automatically generate Boogie programs with
  different characteristics, and test their operational semantics
  against Boogie's .
\item A large experimental evaluation of \bcc,
  which found completeness failures in the latest Boogie version.
\item For reproducibility,
  our implementation of \plt
  and all experimental artifacts are available~\ifblind\textcolor{blue}{[to be released after double-blind review]}\else\cite{BCC-REPP}\fi.
\end{enumerate*}

\section{Related Work}
\label{sec:related-work}

The implementation of a modern formal verification tool
is usually a complex piece of software that integrates
several independently-developed components.
As such, a full end-to-end verification has remained, so far, beyond
the capabilities of verification technology.
The work that is perhaps closest to the ideal of a fully verified verifier
are fully verified \emph{compilers}\iflong
\else~\cite{CompCert1,CompCert2,CakeML1,CakeML2}. \fi 
Interestingly, even a fully verified compiler like CompCert still
benefits from systematic testing~\cite{testingCompCert}.

Similarly to a compiler,
a formal verification tool
usually combines a front-end and a back-end.
It is also increasingly common for verifiers
to use an intermediate verification language---supported by tools like Boogie~\cite{Boogie},
Viper~\cite{Viper}, and Why3~\cite{Why3}---which helps bridge the semantic gap between front-end and back-end.
\iflong

\fi
There exist a few cases of 
formally verified back-end provers~\cite{skotam_creusat_2022,Baek21,FromJ24}
and even intermediate verification condition generators~\cite{HermsMM12}.
Several works prove ``once and for all''
that their front-end \emph{translation} is sound~\cite{VogelsJP09,BackesHT11,SmansJP12,FialaI0PS23},
either just on paper or with a mechanized proof.
However, none of these works extend the soundness proofs
to the actual software \emph{implementations} of the front-end translation.\footnote{
  In a similar vein, there has been work on
  formally proving the correctness of
  verification-condition generation algorithms~\cite{Nipkow98,MartiA09,GreenawayLAK14}---often as part of developing larger verified systems~\cite{KleinAEHCDEEKNSTW10}---in a way that a correct-by-construction implementation
  can be synthesized from the correctness proof.
  While such an approach can produce trustworthy verification components,
  it solves a different problem than verifying an existing implementation.
  For example, a correct-by-construction reimplementation of Boogie
  would not be a perfect replacement unless it offered
  the very same performance, features, and capabilities
  as the actual Boogie tool.
}
A more viable approach than ``once and for all'' verification
is \emph{run validation}:
for any given run of a verifier,
extract a checkable \emph{certificate}
that the translation is sound.
This idea has been applied to validate
Boogie's verification condition generation~\cite{ParthasarathyMS20},
Viper's front-end encoding into Boogie~\cite{ParthasarathyDBMS24},
and verifiers based on the K framework~\cite{LinCTWR23}.

Even in the rare cases where a full formal validation
of a verification tool is feasible,
\emph{lightweight} tools---usually based on testing---offer a different, appealing trade-off:
  while they cannot provide absolute guarantees of correctness,
they are practically useful
to detect bugs,
and to test for properties,
such as completeness/precision, robustness, or scalability,
that are less amenable to formal ``all or nothing'' verification.
In recent years,
techniques based on testing have been applied to
a variety of formal verification tools, including
SMT solvers~\cite{MansurCWZ20,DBLP:conf/kbse/BringolfW022,DBLP:journals/pacmpl/Winterer024},
intermediate verification languages~\cite{boogie-robustness-testing},
software model checkers~\cite{ZhangSYZPS19,FinkBK22,ThobenHW23},
symbolic execution engines~\cite{KapusC17},
and verifiers for reactive systems~\cite{SteffenINMG14}.

Effectively testing a verification tool
requires generating highly structured inputs
and complex ``behavioral'' oracles.
To this end,
it is common to use
grammar-based generation
and
differential testing---two techniques mutuated from
the prolific work on compiler testing~\cite{CSmith,Zhendong-compilertesting,compiler-testing},
which presents similar challenges.
The work presented in this paper 
also employs these techniques
within the \plt framework~\cite{PLT-Redex}:
a ``domain-specific language for semantic models that is embedded in the Racket programming language''~\cite{dsl-redex}.
{}
To our knowledge,
all applications of \plt to date have been to analyze
domain-specific or general-purpose programming languages---often with a functional flavor,
such as JavaScript~\cite{JS-redex} and Lua~\cite{lua-redex}.
One of this paper's contributions
is demonstrating how \plt
is also applicable to test the imperative features
of an intermediate language for verification.

Using a combination of constraint solving and concrete enumeration,
previous work targeted executing the Boogie language,
with the goal of help debug failed verification attempts~\cite{Boogaloo,symbooglix}.
Such a goal is largely complementary to the present paper's:
whereas those tools explore as efficiently as possible a
subset of
the (unbounded) execution space of a given (nondeterministic) Boogie
program,
our goal is checking the consistency of Boogie's
verification semantics with a concrete execution semantics
on a large number of randomly generated Boogie programs.

\section{Modeling a Deterministic Subset of the Boogie Language}
\label{sec:boogie-language-model}

\bcc deploys a formal model
of a deterministic subset
of the Boogie intermediate verification language~\cite{Boogie},
encoded using the PLT Redex framework~\cite{PLT-Redex}.
The \bplsub model consists of three components:
\begin{enumerate*}
\item a \textbf{grammar}
  defining the \textbf{syntax} of \emph{well-formed}
  \bplsub program terms (\autoref{subsec:syntax});

\item an \textbf{operational semantics} consisting of
  reduction rules that specify
  valid executions as transitions between program terms (\autoref{subsec:operational-semantics});

\item \textbf{judgments} consisting of rules
  for name resolution and type checking (\autoref{subsec:judgments}).
\end{enumerate*}

\subsection{Why a \emph{Deterministic} Subset of Boogie?}
\label{sec:why-deterministic}

Before delving into the details of our \plt model of \bplsub,
let's explain why we focused on this specific restricted subset
of the Boogie language.
Boogie combines imperative features
\iflong
\else
  (e.g., variables)
\fi
and specification features
\iflong
\else
  (e.g., preconditions).
\fi
Many of Boogie's features introduce nondeterminism,
which is widely useful for specification,
but is not readily executable.
For example, the statement \Bpl{assume x > 0} is a passive statement
that captures all runs where variable \Bpl{x} is positive.
\iflong
\else
  ``Executing'' the \Bpl{assume} would mean
  either enumerating values of \Bpl{x} up to a finite bound
  or
  defining a symbolic semantics.
\fi
While this different application of \plt could also be interesting,
our present goal is to model Boogie's semantics
in a way that is straightforward and independent of
Boogie's implementation details.
This increases the confidence that our execution model is accurate,
which bolsters its usefulness as an oracle for differential
testing of Boogie.
Therefore, \bplsub
drops all nondeterministic constructs of Boogie.

\subsection{Syntax}
\label{subsec:syntax}

\autoref{fig:bpl0-grammar}
shows the grammar defining the syntax of \bplsub.
Precisely, Boogie's actual syntax is shown on the right,
but in most of the paper we will use \plt's
parenthesized prefix notation (since \plt is implemented in Racket)
shown on the left.

\begin{figure}[t]
  \centering
  \begin{alignat*}{4}
    \text{Program}~P         & ::= (\Rkt{main}\; L\; B)                                                      & \quad & \Bpl{procedure main}()\ \Bpl{returns} ()\;\{L\;B\}                                     \\
    \text{Locals}~L          & ::= \emptyset \mid (\Rkt{let}\ (v\ \Rkt{=}\; \ell\; \Rkt{:}\ t)\; L)          &       & \epsilon \mid \Bpl{var}\ v\colon t := \ell\Bpl{;}\; L                                  \\
    \text{Type}~t            & ::= \Rkt{bool} \mid \Rkt{int}                                                 &       & \Bpl{bool} \mid \Bpl{int}                                                              \\
    \text{Body}~B            & ::= \emptyset \mid (\Rkt{do}\; s\; B)                                         &       & \epsilon \mid s                                                                        \\
    \text{Statement}~s       & ::= (\Rkt{:=}\; v\; e) \mid (\Rkt{if}\; e\; B\; B) \mid (\Rkt{while}\; e\; B) &       & v\; \Bpl{:=}\; e \mid \Bpl{if}\; e\; {B}\; \Bpl{else}\; {B} \mid \Bpl{while}\; e\; {B} \\
                             & \phantom{::=} \mid (\Rkt{assert}\; e)                                         &       & \mid \Bpl{assert}\; e                                                                  \\
    \text{Expression}~e      & ::= \ell \mid v \mid (b\ e\ e) \mid (u\;e)                                    &       & \ell \mid v \mid e\ b\ e \mid u\ e                                                     \\
    \text{Literals}~\ell     & ::= \mathbb{Z} \mid \Rkt{true} \mid \Rkt{false}                               &       &                                                                                        \\
    \text{Binary Operator}~b & ::= \vee \mid \wedge \mid \implies \mid + \mid - \mid / \mid *                &       &                                                                                        \\
    \text{Unary Operator}~u  & ::= \neg \mid -                                                               &       &                                                                                        \\
  \end{alignat*}
  \caption{The syntax of \bplsub in \plt (left),
    and the corresponding Boogie syntax (right).}
  \label{fig:bpl0-grammar}
\end{figure}

A program $P$ consists of a single \Rkt{main} procedure
$(\Rkt{main}\; L\; B)$ with
a local environment $L$
and a body $B$.
A \emph{local environment}
is an inductive list of
\Rkt{let}s $(\Rkt{let}\; (v\; \Rkt{=}\; \ell\; \Rkt{:}\; t))$,
each 
binding a variable symbol $v$
to a literal value $\ell$
of type $t$
(\Rkt{int} or \Rkt{bool}).
A \emph{body}
is an inductive list of statements $s$:
\emph{assignments} $(\Rkt{:=}\ v\ e)$,
\emph{assertions} $(\Rkt{assert}\; e)$,
\emph{conditionals} $(\Rkt{if}\ e\ B_1\ B_2)$,
and \emph{loops} $(\Rkt{while}\; e\; B)$.
The rest of the grammar defines the usual
Boolean and arithmetic \emph{expressions}
used in assignments and conditions.

Even before formally presenting its semantics,
it should be clear that \bplsub is a strictly deterministic language:
\begin{enumerate*}
\item each program has a single entry point to the sole \Rkt{main} procedure;
\item before the body executes,
  the local environment initializes
  every program variable to a literal value;
\item the body's statements execute sequentially;
\item all available statements are deterministic.
\end{enumerate*}

\subsection{Operational Semantics}
\label{subsec:operational-semantics}

\autoref{fig:bpl0-reduction} shows
\bplsub's small-step operational semantics.
Each rule
defines a different case
of the \emph{reduction relation} $A \leadsto B$,
which specifies how
a program term $A$ rewrites into another one $B$
in a way that captures a single evaluation step.

\begin{figure}[t]
  \centering
  \begin{align*}
    E[(b\; \ell_1\; \ell_2)]                                   & \leadsto E[\llbracket b \; \ell_1\; \ell_2 \rrbracket]                                        &  & \text{Binary Evaluation}  \\
    E[(u\; \ell_1)]                                            & \leadsto E[\llbracket u\; \ell_1 \rrbracket]                                                  &  & \text{Unary Evaluation}   \\
    (\Rkt{main}\; L\; \emptyset)                               & \leadsto \Rkt{success}                                                                        &  & \text{Success}            \\
    E[(\Rkt{assert}\; \Rkt{false})]                            & \leadsto \Rkt{failure}                                                                        &  & \text{Failure}            \\
    E[(\Rkt{do}\; (\Rkt{assert}\; \Rkt{true})\; B)]            & \leadsto E[B]                                                                                 &  & \text{Assert True}        \\
    (\Rkt{main}\; L\; E[v])                                    & \leadsto (\Rkt{main}\; L\; E[L\{v\}])                                                         &  & \text{Local Substitution} \\
    (\Rkt{main}\; L\; E[(\Rkt{do}\; (:= v\; \ell)\; B)])       & \leadsto (\Rkt{main}\; L\{v \leftarrow \ell\}\; E[B])                                         &  & \text{Local Assignment}   \\
    E[(\Rkt{do}\; (\Rkt{if}\; \Rkt{true}\; B_1\; B_2)\; B_3)]  & \leadsto E[(\Rkt{do}\; B_1 \cdot B_3)]                                                        &  & \text{If-Then}            \\
    E[(\Rkt{do}\; (\Rkt{if}\; \Rkt{false}\; B_1\; B_2)\; B_3)] & \leadsto E[(\Rkt{do}\; B_2 \cdot B_3)]                                                        &  & \text{If-Else}            \\
    E[(\Rkt{do}\; (\Rkt{while}\; e\; B_1)\; B_2)]              & \leadsto E[(\Rkt{do}\; (\Rkt{if}\; e\; B_1 \cdot (\Rkt{while}\; e\; B_1)\; \emptyset)\; B_2)] &  & \text{Loop}
  \end{align*}
  \caption{Operational semantics of \bplsub using evaluation contexts.}
  \label{fig:bpl0-reduction}
\end{figure}

\begin{figure}[t]
  \centering
  \small
  \[
    \begin{array}{rcl}
      E & ::=  & \Rkt{hole} \mid \Rkt{(} u \; E \Rkt{)} \mid \Rkt{(} b \; E \; e \Rkt{)} \mid \Rkt{(} b \; l \; E \Rkt{)} \\
        & \mid & \Rkt{(:=} \; v \; E \Rkt{)} \mid \Rkt{(assert} \; E \Rkt{)} \mid \Rkt{(if} \; E \; B_1 \; B_2 \Rkt{)}    \\
        & \mid & \Rkt{(do} \; E \; B \Rkt{)} \mid \Rkt{(main} \; L \; E \Rkt{)}
    \end{array}
  \]
  \caption{Specification of the possible evaluation contexts for \autoref{fig:bpl0-reduction}'s rules.}
  \label{fig:bpl0-eval-context}
\end{figure}

Most of the reduction rules in \autoref{fig:bpl0-reduction}
involve an \emph{evaluation context} $E$.
In a nutshell, $E[p]$
means that the program term $p$
can only be reduced (as specified by the rule),
when it is in certain parts of the whole program term.
\autoref{fig:bpl0-eval-context}
lists all possible evaluation contexts,
defined in a way that ensures that the reduction rules
can only be applied in a fixed, unambiguous order,
which matches the program's sequential order.
In particular,
the patterns $E$ are defined in such a way
that the \Rkt{hole} can only match in one position.

\autoref{fig:bpl0-reduction}'s
rules
do not reduce the environment $L$
directly, but use it to keep track
of the current values of the program variables in each step of the reduction:
$L\{v\}$ denotes a lookup of $v$'s current value in $L$,
whereas $L\{v \leftarrow \ell\}$
denotes an environment where $v$ is bound to value $\ell$,
and all other variables are as in $L$.

Most of the meaning of the (other) rules should also be straightforward,
as they simply capture:
\begin{enumerate*}
\item the evaluation of binary and unary operators,
  which are simply reduced to the evaluation $\llbracket \cdot \rrbracket$ of
  the corresponding Racket operators and literals;
\item successful program termination, when all statements have
  been ``consumed'' without errors;
\item program termination with a \Rkt{failure},
  when an \Rkt{assert} evaluates to false;
\item in contrast, an \Rkt{assert} that passes is equivalent to a skip;
\item a conditional reduces to its ``then'' or ``else'' branch
  depending on what Boolean the condition evaluates to;
\item a loop recursively reduces to evaluating its body until
  its condition no longer holds.
\end{enumerate*}

\begin{figure}[!htb]
  \centering
\def\defaultHypSeparation{\hskip 0.4em}
  \begin{minipage}{0.48\textwidth}
\begin{prooftree}
      \LeftLabel{Comp.}
      \AxiomC{$L \vdash e_1:\Rkt{int}$}
      \AxiomC{$L \vdash e_2:\Rkt{int}$}
      \BinaryInfC{$L \vdash (\{<, >, =, \leq, \geq\}\; e_1\; e_2):\Rkt{bool}$}
    \end{prooftree}
    \begin{prooftree}
      \LeftLabel{Bin. \Rkt{int}}
      \AxiomC{$L \vdash e_1:\Rkt{int}$}
      \AxiomC{$L \vdash e_2:\Rkt{int}$}
      \BinaryInfC{$L \vdash (\{+, -, *, /\}\; e_1\; e_2):\Rkt{int}$}
    \end{prooftree}
    \begin{prooftree}
      \LeftLabel{Bin. \Rkt{bool}}
      \AxiomC{$L \vdash e_1:\Rkt{bool}$}
      \AxiomC{$L \vdash e_2:\Rkt{bool}$}
      \BinaryInfC{$L \vdash (\{\wedge,\vee,\implies,=\}\; e_1\; e_2):\Rkt{bool}$}
    \end{prooftree}
    \begin{prooftree}
      \LeftLabel{Un. \Rkt{int}}
      \AxiomC{$L \vdash e:\Rkt{int}$}
      \UnaryInfC{$L \vdash (-\; e):\Rkt{int}$}
    \end{prooftree}
    \begin{prooftree}
      \LeftLabel{Un. \Rkt{bool}}
      \AxiomC{$L \vdash e:\Rkt{bool}$}
      \UnaryInfC{$L \vdash (\neg\; e):\Rkt{bool}$}
    \end{prooftree}
    \begin{prooftree}
      \LeftLabel{Vars}
      \AxiomC{$(v:t) \in L$}
      \UnaryInfC{$L \vdash v:t$}
    \end{prooftree}
  \end{minipage}
  \begin{minipage}{0.48\textwidth}
\begin{prooftree}
      \LeftLabel{Assign}
      \AxiomC{$L \vdash v:t$}
      \AxiomC{$L \vdash e:t$}
      \AxiomC{$L \vdash B$}
      \TrinaryInfC{$L \vdash (\Rkt{do}\; ( \Rkt{:=}\; v\; e)\; B)$}
    \end{prooftree}
    \begin{prooftree}
      \LeftLabel{Assert}
      \AxiomC{$L \vdash e:\Rkt{bool}$}
      \AxiomC{$L \vdash B$}
      \BinaryInfC{$L \vdash ({\Rkt{do}\; (\Rkt{assert}}\; e)\; B)$}
    \end{prooftree}
    \begin{prooftree}
      \LeftLabel{If}
      \AxiomC{$L \vdash e:\Rkt{bool}$}
      \AxiomC{$L \vdash B_1, B_2, B_3$}
      \BinaryInfC{$L \vdash (\Rkt{do}\; (\Rkt{if}\; e\; B_1\; B_2)\; B_3)$}
    \end{prooftree}
    \begin{prooftree}
      \LeftLabel{While}
      \AxiomC{$L \vdash e:\text{bool}$}
      \AxiomC{$L \vdash B_1, B_2$}
      \BinaryInfC{$L \vdash (\Rkt{do}\; ( \Rkt{while}\; e\; B_1 )\; B_2)$}
    \end{prooftree}

\begin{prooftree}
      \LeftLabel{Empty program}
      \AxiomC{}
      \UnaryInfC{$L \vdash \emptyset$}
    \end{prooftree}
    \begin{prooftree}
      \LeftLabel{Env.}
      \AxiomC{$\emptyset \vdash \ell:t$}
      \AxiomC{$v \notin L$}
      \AxiomC{$\vdash L$}
      \TrinaryInfC{$\vdash (\Rkt{let}\; [v=\ell:t]\; L)$}
    \end{prooftree}
    \begin{prooftree}
      \LeftLabel{Main}
      \AxiomC{$\vdash L$}
      \AxiomC{$L \vdash B$}
      \BinaryInfC{$\vdash (\Rkt{main}\; L\; B)$}
    \end{prooftree}
  \end{minipage}
  \caption{Inference rules for \bplsub's type system in the \plt model.}
  \label{fig:bpl0-type-system}
\end{figure}

\subsection{Judgments}
\label{subsec:judgments}

\bcc can generate \emph{well-formed} \bplsub programs
using \autoref{fig:bpl0-grammar}'s grammar.
A well-formed program may be invalid
because the grammar alone does not ensure
that the program is also \emph{well-named}
(variables are declared and initialized before their first usage)
and \emph{well-typed}
(expression values conform to their expected \Rkt{int} or \Rkt{bool} type).
\bcc models \bplsub's name resolution and typing rules
using \plt's judgments~\cite{FetscherCPHF15},
which can be used by the generation algorithm
to automatically produce \bplsub programs that are
well-named and/or well-typed.

\autoref{fig:bpl0-type-system} presents
\bplsub's type system, where
$L \vdash e \colon t$ denotes, as customary,
that term $e$ has type $t$ when evaluated within local environment $L$,
whereas $L \vdash s$ means that statement $s$ is well-typed.
Notice that rules \emph{Vars} and \emph{Env.}
also determine whether a program is \emph{well-named}
by checking that every used variable was declared and initialized (\emph{Vars}),
and that no variable is declared twice (\emph{Env.}).
In practice, \bcc includes two sets of judgments:
the one in \autoref{fig:bpl0-type-system}
checks well-typedness as well as well-namedness;
and another one that \emph{only} checks well-namedness.

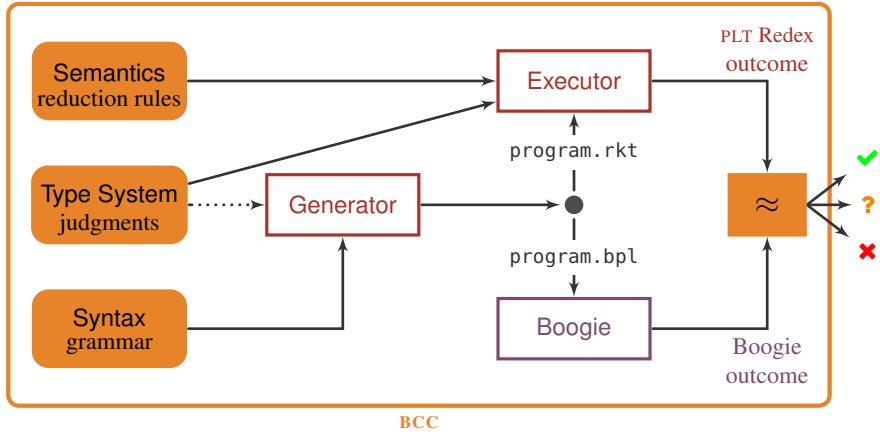
\begin{figure}[!htb]
\centering
\begin{tikzpicture}[
  databox/.style={rectangle,very thick,
    rounded corners=2mm,font=\footnotesize\sffamily,
    minimum width=20mm,minimum height=10mm,
    draw=bcccol,fill=bcccol,text=black,
    text height=0mm,
    label={[below=15pt]#1}},
  toolbox/.style={rectangle,very thick,
    font=\footnotesize\sffamily,
    minimum width=20mm,minimum height=8mm,
    draw=#1,text=#1},
  bcctoolbox/.style={rectangle,very thick,
    minimum width=10mm,minimum height=8mm,
    font=\footnotesize\sffamily,
    draw=bcccol,fill=bcccol,text=black},
  align=center,
  node distance=12mm and 20mm,
  ]

  \matrix[row sep=6mm,column sep=10mm] {
    \node[databox={reduction rules}] (semantics) {Semantics};
    &
    & \node[toolbox={pltcol}] (executor) {Executor};
    & \coordinate (check-above);
    \\
    \node[databox={judgments}] (judgments) {Type System};
    & \node[toolbox={pltcol}] (generator) {Generator};
    & \node[circle, fill=black!70, minimum size=7pt, outer sep=1.5pt, inner sep=2pt] (program) {};
    & \node[bcctoolbox] (check) {\large$\approx$};
    \\
    \node[databox={grammar}] (grammar) {Syntax};
    &
    &
    \node[toolbox={boogiecol}] (boogie) {Boogie};
    & \coordinate (check-below);
    \\
  };

  \coordinate[right=8 mm of check] (outcome);
  \node  (unknown) at (outcome) {\outUN};
  \node[above=4mm of outcome] (ok) {\outOK};
  \node[below=4mm of outcome] (no) {\outNO};

  \node [
  fit=(semantics)(grammar)(judgments)(generator)(executor)(boogie)(check),
  draw=bcccol,ultra thick,rounded corners,inner ysep=5mm,inner xsep=3mm,
  label={[bcccol]south:\textbf{\bcc}}
  ] (bcc) {};

  \begin{scope}[color=black!80,line width=1pt,round cap-latex',every node/.style={font=\footnotesize}]
    \draw (semantics) -- (executor);
    \draw (grammar.east) -| (generator.south);
    \draw[dotted] (judgments) -- (generator);
    \draw (judgments) -- (executor);
    \draw (generator) -- (program);
    \draw (program) -- node [fill=white] {\texttt{program.rkt}} (executor);
    \draw (program) -- node [fill=white] {\texttt{program.bpl}} (boogie);
    \draw (executor) -| node[above,align=center,text=pltcol] {\plt \\ outcome} (check);
    \draw (boogie) -| node[below,align=center,text=boogiecol] {Boogie\\outcome} (check);
\draw (check.east) -- (ok);
    \draw (check.east) -- (no);
    \draw (check.east) -- (unknown);
  \end{scope}

\end{tikzpicture}
\caption{An overview of how \bcc works.}
\label{fig:workflow}
\end{figure}

\section{How \bcc Works}
\label{sec:methodology}

\autoref{fig:workflow}
overviews how \bcc uses the \plt model of \bplsub,
described in \autoref{sec:boogie-language-model},
to:
\begin{enumerate*}
\item \emph{Generate} a large number of
  \bplsub programs;
\item \emph{Execute} them according to their operational semantics;
\item \emph{Verify} them using the Boogie verifier;
\item \emph{Compare} the outcome of execution and verification,
  reporting any inconsistency that may be indicative of a
  failure.
\end{enumerate*}
The following sections describe each step of \bcc's analysis.

\subsection{Generator}
\label{subsec:program-generation}

\bcc uses \plt's term generation capabilities to produce
a large number of \bplsub programs.
As explained in \autoref{subsec:judgments},
\plt can produce three kinds of programs:
\begin{enumerate*}
\item \emph{well-formed} programs, which conform to \bplsub's grammar;
\item \emph{well-named} programs, which are well-formed
  and also free from any usage of uninitialized variables;
\item \emph{well-typed} programs, which
  are well-named and also typecheck correctly.
\end{enumerate*}
Each kind of program is suitable to test different
aspects of Boogie: its verification process (well-typed),
its typechecker (well-named),
or its name resolver (well-formed).\footnote{
  In this work, we did not experiment with generating
  syntactically incorrect Boogie programs, which would
  only test Boogie's parser. To this end, there are a number of other techniques, such as fuzzing,
  that do not require a formal model of Boogie's type system and semantics.
}

\bcc's generator actually produces each program in two forms:
one uses Racket syntax, and is used by the \plt \emph{Executor};
the other one uses Boogie's concrete syntax, and is fed to \emph{Boogie} itself.

\begin{table}[!htb]
\centering
\renewcommand{\arraystretch}{0.2}
\begin{tabular}{m{0.35\textwidth} m{0.3\textwidth} m{0.3\textwidth}}
\begin{minipage}[t]{\textwidth}
\begin{lstlisting}[language=boogie]
procedure success() {
  var x: int := 0;
  assert x == 0;
}
\end{lstlisting}
\end{minipage}
&
\begin{minipage}[t]{\textwidth}
\begin{lstlisting}[language=boogie]
procedure failure() {
  var x: int := 3;
  assert x < 0;
}
\end{lstlisting}
\end{minipage}
&
\begin{minipage}[t]{\textwidth}
\begin{lstlisting}[language=boogie]
procedure name_error() {
  var x: int := 0;
  assert y == 0;
}
\end{lstlisting}
\end{minipage}
\\

\begin{minipage}[t]{\textwidth}
\begin{lstlisting}[language=boogie]
procedure type_error() {
  var x: int := 0;
  assert x;
}
\end{lstlisting}
\end{minipage}
&
\begin{minipage}[t]{\textwidth}
\begin{lstlisting}[language=boogie]
procedure loop() {
  var x: int := 3;
  while (true) {
    x := 0;
  }
  assert false;
}
\end{lstlisting}
\end{minipage}
&
\begin{minipage}[t]{\textwidth}
\begin{lstlisting}[language=boogie]
procedure timeout() {
  var x: int := 0;
  while (x < 1000000) {
    x := x + 1;
  }
  assert false;
}
\end{lstlisting}
\end{minipage}
\end{tabular}
\caption{Examples of Boogie programs with various execution and verification outcomes.}
\label{tab:boogie-outcome-programs}
\end{table}

\subsection{Executor}
\label{subsec:program-execution}

Given a \bplsub program $P$,
\bcc first uses the model's \emph{judgments}
to check whether $P$ is
well-named
and well-typed.
Obviously, these checks are only necessary
if $P$ was not already generated in a well-typed form by construction.
In principle, \bcc
could execute $P$ without performing these checks beforehand,
so that any name resolution or type error would result in a runtime failure.
However, performing name resolution and type checking
\emph{before} executing a program
is consistent with Boogie's behavior:
it does not run $P$ unless it typechecks.
In this way, \bcc's execution of $P$
is directly comparable to Boogie's verification run on $P$.

If $P$ the judgment checks fail,
\bcc reports one of the following outcomes:
\begin{description}
\item[\outcome{name-error}:] $P$ is not well-named; for
  example, program \Bpl{name_error} in \autoref{tab:boogie-outcome-programs}
  results in a \outcome{name-error} because variable \Bpl{y}
  is used but not declared.
\item[\outcome{type-error}:] $P$ is not well-typed; for
  example, program \Bpl{type_error} in \autoref{tab:boogie-outcome-programs}
  results in a \outcome{type-error} because \Bpl{int} variable \Bpl{x}
  cannot be used as a \Bpl{bool} assertion.
\end{description}
If $P$ is well-typed, \bcc \emph{executes} it by
repeatedly applying \autoref{fig:bpl0-reduction}'s reduction rules
until one of the following outcomes is reached:
\begin{description}
\item[\outcome{success}:] $P$ evaluates to an empty body $\emptyset$,
  which means that the program executed and terminated without any errors;
  for example, program \Bpl{success} in \autoref{tab:boogie-outcome-programs}
  clearly terminates with a \outcome{success}.
\item[\outcome{failure}:] $P$ evaluates to \Rkt{(assert false)},
  which means that the program triggered an assertion failure at some point
  during its execution;
  for example, program \Bpl{failure} in \autoref{tab:boogie-outcome-programs}
  ends with an assertion \outcome{failure} because $\Bpl{x} = 3 > 0$.
\item[\outcome{loop}:] $P$ evaluates to a term $T$,
  which is identical to a term produced during a previous step of the
  evaluation; this means that the program entered an infinite loop.
  For example, program \Bpl{loop} in \autoref{tab:boogie-outcome-programs}
  results in an infinite loop where the program state does not change.
  Programs that result in outcome \outcome{loop}
  are still (partially) correct, in that they do not trigger assertion failures:
  the \Bpl{assert false} in program \Bpl{loop} never gets
  executed because it is effectively unreachable.
\item[\outcome{timeout:}] $P$'s evaluation reaches a maximum number of steps
  (by default, \numprint{100000}) and is forcefully terminated;
  for example, program \Bpl{timeout} in \autoref{tab:boogie-outcome-programs}
  exceeds the default maximum number of steps,
  and hence it would be terminated by \bcc.
  As usual, a \outcome{timeout} is inconclusive as to whether $P$
  is correct or not: program \Bpl{timeout} would eventually trigger
  an assertion failure if its loop were run to completion;
  but a variant of \outcome{timeout} with loop condition \Bpl{true}
  would instead never terminate (\Bpl{int}s are unbounded in Boogie).
\end{description}

\subsection{Boogie}
\label{subsec:program-verification}

\bcc also feeds each generated program $P$ to the Boogie verifier,
which first checks whether $P$ is well-named and well-typed;
if it is, Boogie performs a deductive static verification of $P$
by expressing its correctness conditions as logic formulas,
whose validity is checked by the automated theorem prover Z3.

Thus, a Boogie verification run on $P$ reports one of the following outcomes:
\begin{description}
\item[\outcome{name-error}:] $P$ fails name resolution; for example program \Bpl{name_error} in \autoref{tab:boogie-outcome-programs}.
\item[\outcome{type-error}:] $P$ fails type checking; for example program \Bpl{type_error} in \autoref{tab:boogie-outcome-programs}.
\item[\outcome{success}:] $P$ verifies correctly, which means that
  all possible executions of $P$ are free from assertion failures;
  for example, program \Bpl{success} in \autoref{tab:boogie-outcome-programs}.
  Boogie does not check termination of loops, but only partial correctness;
  therefore, program \Bpl{loop} leads to a \outcome{success} in Boogie,
  because it recognizes that the \Bpl{assert false} is effectively unreachable.
\item[\outcome{failure}:] $P$ fails verification,
  which means that some assertions in $P$ may not hold;
  for example, program \Bpl{failure} in \autoref{tab:boogie-outcome-programs}.
\item[\outcome{timeout}:] Boogie's verification run on $P$ 
  is forcefully terminated after it does not return within a timeout
  (by default, one hour). Boogie's timeouts are quite rare on the
  kinds of programs generated by \bcc
  (which usually verify in a matter of seconds),
  but sometimes Boogie loses control of the Z3 process it started
  and does not return within a reasonable amount of time.
  As usual, we treat a \outcome{timeout} as an inconclusive verification result.
\end{description}

Outcome \outcome{loop} 
is inapplicable to Boogie's verification:
since it does not analyze termination,
it won't report if a loop iterates forever.
Finally, a Boogie verification run may also result in
a parsing error.
Since all the programs generated by \bcc are
(at least) syntactically correct,
this outcome is irrelevant for our work.

\begin{table}[!tbh]
  \centering
  \setlength{\tabcolsep}{8pt}
  \begin{tabular}{lccccc}
    \toprule
    \multicolumn{1}{c|}{} & \multicolumn{5}{c}{\textbf{Boogie outcome $b$}} \\
    \multicolumn{1}{c|}{\textbf{\plt outcome $p$}} & \outcome{success} & \outcome{failure} & \outcome{timeout} & \outcome{name-error} & \outcome{type-error} \\
    \midrule
    \outcome{success}
     & \outOK        & \outNO   & \outUN   & \outNO   & \outNO           \\
    \outcome{failure}
    & \outNO         & \outOK   & \outUN    & \outNO  & \outNO           \\
    \outcome{loop}
    & \outOK        & \outNO    & \outUN   & \outNO   & \outNO           \\
    \outcome{timeout}
    & \outUN        & \outUN    & \outUN   & \outNO   & \outNO           \\
    \outcome{name-error}
    & \outNO        & \outNO    & \outNO   & \outOK   & \outNO           \\
    \outcome{type-error}
    & \outNO        & \outNO    & \outNO   & \outNO   & \outOK           \\
    \bottomrule
  \end{tabular}
  \caption{Consistency between \plt execution and Boogie verification outcomes.}
  \label{tab:outcome-match}
\end{table}

\subsection{Consistency Checking}
\label{subsec:checking-consistency}

Finally, \bcc compares the outcome of
\plt execution according to the operational semantics
and Boogie verification for the same program $P$,
and reports any \emph{inconsistency}.
This step corresponds to box $\approx$ in \autoref{fig:workflow}.

\autoref{tab:outcome-match}
shows what \bcc's consistency check reports
for each combination of \plt outcome $p$ and Boogie outcome $b$.
\begin{description}
\item[\outOK] denotes that $p$ and $b$ are consistent; hence,
  we have successfully tested Boogie's correct behavior on a program $P$.
\item[\outNO] denotes that $p$ and $b$ are inconsistent;
  hence, we have exposed a failure in Boogie's behavior,
  which is inconsistent with the expected semantics of $P$.
\item[\outUN] denotes an inconclusive test: $p$ and $b$
  are not directly comparable, and hence Boogie's behavior may
  or may not be correct.
\end{description}

As shown in \autoref{tab:outcome-match},
if $p$ and $b$ are literally the same conclusive outcome
(\outcome{success} or \outcome{failure})
then the consistency check obviously results in \outOK.
However, also the combination $p = \outcome{loop}$ and $b = \outcome{success}$
is consistent (\outOK):
since \outcome{loop} denotes a provably infinite loop,
Boogie should also be able to verify that the program in question is (partially) correct,
in that any code up to and until the loop is free from errors,
whereas the code after the loop is unreachable,
and hence irrelevant for correctness.

In contrast, if $p = \outcome{failure}$ but $b = \outcome{success}$,
we would conclude that \bcc has found a \emph{soundness} failure \outNO,
since Boogie erroneously passed as correct an incorrect program
(a false negative).
Conversely,
if $p = \outcome{success}$ but $b = \outcome{failure}$,
we would conclude that \bcc has found a \emph{completeness} failure \outNO,
since Boogie reported a spurious verification failure for a correct program
(a false positive);
as explained above, the same conclusion \outNO\ 
holds if $p = \outcome{loop}$ but $b = \outcome{failure}$.

Clearly,
if $p = \outcome{name-error}$---that is, the program is not name-correct---Boogie should report the same outcome, in which case \bcc reports consistency \outOK;
any other Boogie outcome would result in \outNO,
and reveal that Boogie's name resolution pass has a bug.
Similarly,
if $p = \outcome{type-error}$---that is, the program is not type-correct---Boogie should report the same outcome,
unless its type checker has a bug.

Finally, if $p = \outcome{timeout}$,
the consistency check is inconclusive (\outUN) regardless of whether
$b = \outcome{success}$, $b = \outcome{failure}$, or $b = \outcome{timeout}$;
similarly, if $b = \outcome{timeout}$,
the consistency check is inconclusive (\outUN) regardless of whether
$p = \outcome{success}$, $p = \outcome{failure}$, or $p = \outcome{loop}$.
For example, as discussed previously,
program \Bpl{timeout} in \autoref{tab:boogie-outcome-programs}
results in $p = \outcome{timeout}$ and $b = \outcome{failure}$,
which is consistent because the program is actually incorrect;
however, if we removed the \Bpl{assert false},
\plt's outcome $p$ would still be \outcome{timeout}
but Boogie would correctly report $b = \outcome{success}$ because
the program would now be correct.

\begin{figure}[!h]
\centering
\begin{subfigure}[T]{0.33\linewidth}
    \centering
\begin{lstlisting}[language=boogie]
while ($C$)
  invariant $J$ {
  // ...
}
assert $A$;
\end{lstlisting}
\end{subfigure}
\begin{subfigure}[T]{0.33\linewidth}
    \centering
\begin{lstlisting}[language=boogie]
while ($C$)
  invariant $J$ {
  assert $A$;
  // ...
}
\end{lstlisting}
\end{subfigure}
\caption{Generic Boogie loops annotated with loop invariants.}
\label{fig:loop-with-inv}
\end{figure}

\paragraph{Loop invariants.}
To properly gauge the inconsistencies in our experiments that point to
\emph{completeness} failures of Boogie, we need to understand the role played
by loop \emph{invariants}.
Consider \autoref{fig:loop-with-inv}'s
generic loops (annotated with an invariant $J$),
and suppose that 
Boogie's outcome $b$ is a \outcome{failure} of \Bpl{assert $\:A$},
whereas \plt reports $p = \outcome{loop}$ (program on the left)
or $p = \outcome{success}$ (program on the right)
on the same program---hence, the program is (partially) correct.
For the program on the left,
if $\neg C \Rightarrow A$ holds
we say that it is a case of \emph{reasoning} incompleteness of Boogie;
otherwise, we call it 
a case of \emph{annotation} incompleteness.
Similarly, for the program on the right,
if $\neg C$ holds at the loop's entrance
we say that it is a case of \emph{reasoning} incompleteness;
otherwise, we call it a case of \emph{annotation} incompleteness.
In principle, annotation incompleteness
merely indicates that the user should have annotated
the loop with an invariant strong enough to characterize
the loop's behavior:
if $J$ is indeed a correct invariant of the loop,
then Boogie could use it to prove that
$J \land \neg C \Rightarrow A$ (program on the left)
or $J \Rightarrow \neg C$ (program on the right),
hence establishing that the program is correct after all.
Conversely, reasoning incompleteness means
that $A$ holds independent of any loop invariants,
because it is a direct consequence of the loop's condition $C$;
hence, Boogie should be able to successfully verify the program without
additional annotations.
These definitions generalize to cases where the assertion
that spuriously fails in Boogie is not
immediately after the loop (resp.~at the beginning of the loop body)
but follows a sequence $S$ of statements.

Distinguishing between the two kinds of incompleteness
is undecidable in general, because it requires a sound and complete
loop invariant inference.
This also explains why \bplsub does not include loop invariant annotations:
synthesizing correct (and useful) loop invariants for randomly generated
loops is nontrivial, and would require to introduce different techniques.
\autoref{sec:results-consistency} describes how we
tried to estimate annotation vs.\ reasoning incompleteness in our experiments.

\section{Experiments}
\label{sec:experiments}

To evaluate \bcc's capabilities,
we used it to generate a large number of \bplsub programs
with different characteristics,
and to compare the outcome of \plt execution and Boogie verification
on those programs.

\subsection{Setup}
\label{subsec:setup}

\paragraph{Batches.}
Each \emph{batch} of generated programs
is defined by three parameters:
\begin{enumerate*}
\item the \emph{number} of \bplsub programs in the batch;
\item whether they should be well-\emph{typed}, well-\emph{named}, or simply well-\emph{formed};
\item the maximum \emph{size} of each generated program.\footnote{
    \plt's documentation is somewhat vague
    about how this \Rkt{size-expr} parameter is used,
    but it's probably an upper bound on the maximum depth
    of the generated term tree.
    }
\end{enumerate*}
In our experiments, we generated a total of 3 million
syntactically different programs in 12 batches:
for each kind of well-typed, well-named, and well-formed programs,
\begin{enumerate*}
\item \numprint{100000} programs with maximum size 3,
\item \numprint{200000} programs with maximum size 5,
\item \numprint{200000} programs with maximum size 7,
\item \numprint{500000} programs with maximum size 10.
\end{enumerate*}
We chose these numbers after some informal experiments
with \bcc, which suggested two guidelines:
first, even though a larger maximum program size should permit
also small programs, using batches with different maximum sizes
achieves a better program size diversity;
second, batches with a smaller maximum program size
should be smaller, otherwise there is a risk of inefficiently generating
a lot of nearly identical programs.

\paragraph{System.}
The experiments ran on a virtual machine
with 64 cores of an AMD Epyc~7713 3.7~Ghz processor
with 128 GB of RAM, running Ubuntu 22.04,
Racket/\plt~v.~8.2,
Boogie v.~3.4.3
with Z3 v.~4.8.8---the latest versions of the software at the time of these experiments.
To speed up the experiments,
several instances of \bcc ran in parallel processes,
each on different \bplsub programs
and coordinated by a work-stealing algorithm;
precisely, the generation phase used 64 parallel processes,
the execution phase 128, and
the verification phase 512.

\subsection{Results: Generation, Execution, and Verification}
\label{subsec:results-gen-exec-verif}

\begin{table}[!tb]
\centering
\scriptsize
\setlength{\tabcolsep}{2.4pt}
\begin{tabular}{ll *{6}{rrr}}
\toprule
  \multicolumn{1}{c}{\textbf{Batch}} & \multicolumn{1}{c|}{\textbf{Size}} & \multicolumn{3}{c|}{\textbf{Locals $L$}} & \multicolumn{3}{c|}{\textbf{Statement $s$}} & \multicolumn{3}{c|}{\textbf{Expr $e$ (arith)}} & \multicolumn{3}{c|}{\textbf{Expr $e$ (bool)}} & \multicolumn{3}{c|}{\textbf{Expr $e$ (comp)}} & \multicolumn{3}{c}{\textbf{Literals $l$}}
                                                                                                                                                                                     \\
                               & 
\multicolumn{1}{c|}{}          & 
 \multicolumn{1}{c}{min} & \multicolumn{1}{c}{med} & \multicolumn{1}{c|}{max} &
 \multicolumn{1}{c}{min} & \multicolumn{1}{c}{med} & \multicolumn{1}{c|}{max} &
 \multicolumn{1}{c}{min} & \multicolumn{1}{c}{med} & \multicolumn{1}{c|}{max} &
                                                                                \multicolumn{1}{c}{min} & \multicolumn{1}{c}{med} & \multicolumn{1}{c|}{max} &                                                                                    \multicolumn{1}{c}{min} & \multicolumn{1}{c}{med} & \multicolumn{1}{c|}{max} &                                                                                    \multicolumn{1}{c}{min} & \multicolumn{1}{c}{med} & \multicolumn{1}{c}{max}                    \\
  \midrule
  \multirow{4}{*}{\textsl{formed}} & \termrow{well-formed}{3} \\
                               & \termrow{well-formed}{5} \\
                               & \termrow{well-formed}{7} \\
                               & \termrow{well-formed}{10} \\
  \cmidrule(l){1-20}
\multirow{4}{*}{\textsl{named}}
                               & \termrow{well-named}{3} \\
                               & \termrow{well-named}{5} \\
                               & \termrow{well-named}{7} \\
                               & \termrow{well-named}{10} \\
  \cmidrule(l){1-20}
\multirow{4}{*}{\textsl{typed}}
                               & \termrow{well-typed}{3} \\
                               & \termrow{well-typed}{5} \\
                               & \termrow{well-typed}{7} \\
                               & \termrow{well-typed}{10} \\
  \bottomrule
\end{tabular}
\caption{Statistics about the composition
  of the \bplsub programs generated in the experiments.
  For each batch of \emph{well-formed}, \emph{well-named}, or \emph{well-typed}
  programs of a given maximum \emph{size},
  the table reports \emph{min}imum, \emph{med}ian, and \emph{max}imum number of
terms of each kind in any program in the batch.}
\label{tab:term-distributions}
\end{table}

\subsubsection{Generation}
In our experiments, \bcc took about 30 minutes to generate the 3~million programs
in 12 batches.
\autoref{tab:term-distributions}
shows basic statistics about the actual size of the \bplsub programs
generated in each batch,
measured in terms of 
the number of local variables, statements, arithmetic, Boolean and comparison operators, and literals that appear in a program.
Most programs have a non-trivial size in terms of
statements and expressions, whereas they tend to include few
local variables (median: 1--2, maximum: 10).
Well-typed programs are about as large, on average,
as well-named and well-formed programs, but they 
tend to include larger expressions.
This is an indirect effect of using typing judgments as a constraint
to generate \emph{well-typed} programs:
this additional condition
nudges the generation to produce more ``interesting'', deeply nested
expressions;
in contrast, the generation of well-formed programs
has fewer constraints,
and hence results in a more shallow enumeration
of all possible literal/operator combination.
Overall, \bcc managed to generate a broad range of programs
with different sizes and characteristics.

\begin{table}[!tb]
\def\NAN{0}
\centering
\scriptsize
\setlength{\tabcolsep}{1pt}
\begin{adjustwidth}{-2pt}{0mm}
\begin{tabular}{ll *{6}{r} | *{5}{r}}
\toprule
\textbf{Batch} & \textbf{Size} & \multicolumn{6}{c|}{\textbf{\plt Execution Outcomes}} & \multicolumn{5}{c}{\textbf{Boogie Verification Outcomes}} \\
& & \multicolumn{1}{c}{\outcome{success}} & \multicolumn{1}{c}{\outcome{failure}} & \multicolumn{1}{c}{\outcome{loop}} & \multicolumn{1}{c}{\outcome{timeout}} & \multicolumn{1}{c}{\outcome{name-error}} & \multicolumn{1}{c|}{\outcome{type-error}} & \multicolumn{1}{c}{\outcome{success}} & \multicolumn{1}{c}{\outcome{failure}} & \multicolumn{1}{c}{\outcome{timeout}} & \multicolumn{1}{c}{\outcome{name-error}} & \multicolumn{1}{c}{\outcome{type-error}} \\
\midrule
  \multirow{4}{*}{formed}
               & \execVerifRow{well-formed}{3} \\
               & \execVerifRow{well-formed}{5} \\
               & \execVerifRow{well-formed}{7} \\
               & \execVerifRow{well-formed}{10} \\
\cmidrule(){2-8}
\cmidrule(l){9-13}
  \multirow{2}{*}{\textsl{all}}
               & \execVerifGroupRow{well-formed}
\midrule
\multirow{4}{*}{named} 
               & \execVerifRow{well-named}{3} \\
               & \execVerifRow{well-named}{5} \\
               & \execVerifRow{well-named}{7} \\
               & \execVerifRow{well-named}{10} \\
\cmidrule(){2-8}
\cmidrule(l){9-13}
  \multirow{2}{*}{\textsl{all}}
               & \execVerifGroupRow{well-named}
\midrule
\multirow{4}{*}{typed} 
               & \execVerifRow{well-typed}{3} \\
               & \execVerifRow{well-typed}{5} \\
               & \execVerifRow{well-typed}{7} \\
               & \execVerifRow{well-typed}{10} \\
\cmidrule(){2-8}
\cmidrule(l){9-13}
  \multirow{2}{*}{\textsl{all}}
               & \execVerifGroupRow{well-typed}
\midrule
\multirow{2}{*}{\textsl{all}}
 & \execVerifTotalRow
\bottomrule
\end{tabular}
\end{adjustwidth}
\caption{Statistics about the outcome of \emph{\plt execution} (left)
  and \emph{Boogie verification} (right)
  of the \bplsub programs generated in the experiments.
  For each batch of \emph{well-formed}, \emph{well-named}, or \emph{well-typed}
  programs of a given maximum \emph{size},
  the number of programs in the batch
  whose \plt execution resulted in a certain outcome.
  Rows \emph{all} aggregate
  the counts \# and percentage \% within each batch, and in all dataset.}
\label{tab:execution-verification-outcomes}
\end{table}

\subsubsection{Execution}
In our experiments, \bcc took around 110 minutes to execute the 3~million programs
according to \bplsub's operational semantics.
\autoref{tab:execution-verification-outcomes}'s left half
shows the number of programs in each batch whose \plt execution resulted
in one of five possible outcomes: \outcome{success}, \outcome{failure},
\outcome{loop}, \outcome{timeout}, \outcome{name-error}, and \outcome{type-error}.
As expected, most well-formed programs resulted in a name resolution error;
and most well-named programs in a type checking error
(even though a few managed to pass typechecking, and even to pass execution).

As for well-typed programs,
less than \n:{1}|
of them timed out; thus,
most executions ran to completion.
Similarly, only about \n[0]{Execution-Outcomes/well-typed-success/Percentage}|
of the programs terminated with a \outcome{success};
this indicates that
it is more likely that a randomly generated assertion
fails than it holds.
In fact, \outcome{failure} was by far the most common outcome
of executing well-typed programs: over $2/3$ of them terminated
with an assertion failure.
Within each batch of a certain maximum size,
the percentage of programs that fail is similar (around \n:[0]{.65100}[100]|)
but becomes a bit higher for the batch with size 10 (around \n:[0]{.6833480}[100]|),
arguably 
because larger random expression are more likely to introduce
a contradiction.
Nevertheless, \plt detected an infinite \outcome{loop}
in nearly $1/3$ of the randomly generated programs.
In all, despite the simplicity of the Boogie \bplsub subset it targets,
the programs generated by \bcc have a good variety
of possible outcomes, and thus have the potential of exercising the verifier
in different conditions.

\subsubsection{Verification}
In our experiments, \bcc took around 12 hours to verify the 3~million programs
with Boogie;
unsurprisingly, running Boogie is an order of magnitude more time consuming
than running \plt on relatively small programs (or generating them),
since each Boogie run usually involves calls to an SMT solver.
As during \plt execution,
Boogie reports a \outcome{name-error} (resp.~\outcome{type-error}),
on most well-formed (resp.~well-named) programs.

The distribution of verification outcomes for
well-typed programs may seem quite different from
that of execution outcomes, but it is actually largely consistent.
In fact, remember that both execution outcomes $p = \outcome{success}, \outcome{loop}$ correspond to the same verification outcome $b = \outcome{success}$---since a program with an infinite loop should be classified as
correct by Boogie, which only checks partial correctness.
Boogie timeouts are exceedingly rare,
at least with the kinds of programs that \bcc
generates, which do not include complex quantified formulas
that may trip up the quantifier instantiation algorithms~\cite{stabilizing-triggers,CF-iFM17}.
However, Boogie timeouts do occasionally, and unpredictably, happen
on programs that, upon inspection, do not seem to have any
complex or unusual feature that stands out.

\begin{table}[!tbh]
  \def\NAN{}
  \centering
  \setlength{\tabcolsep}{2pt}
  \scriptsize
  \begin{tabular}{l *{5}{rr}}
    \toprule
    \multicolumn{1}{c|}{}                          & \multicolumn{10}{c}{\textbf{Boogie outcome $b$}}                                                                                                                     \\
    \multicolumn{1}{c|}{\textbf{\plt outcome $p$}} & \multicolumn{2}{c|}{\outcome{success}}             & \multicolumn{2}{c|}{\outcome{failure}}             & \multicolumn{2}{c|}{\outcome{timeout}}             & \multicolumn{2}{c|}{\outcome{name-error}}             & \multicolumn{2}{c}{\outcome{type-error}}             \\
    \midrule
    \outcome{success}                              & \bvpcell{success}{success}    & \bvpcell*{success}{failure}    & \bvpcell{success}{other}    & \bvpcell{success}{name-error}    & \bvpcell{success}{type-error}    \\
    \outcome{failure}                              & \bvpcell{failure}{success}    & \bvpcell{failure}{failure}    & \bvpcell{failure}{other}    & \bvpcell{failure}{name-error}    & \bvpcell{failure}{type-error}    \\
    \outcome{loop}                                 & \bvpcell{loop}{success}       & \bvpcell*{loop}{failure}       & \bvpcell{loop}{other}       & \bvpcell{loop}{name-error}       & \bvpcell{loop}{type-error}       \\
    \outcome{timeout}                              & \bvpcell{timeout}{success}    & \bvpcell{timeout}{failure}    & \bvpcell{timeout}{other}    & \bvpcell{timeout}{name-error}    & \bvpcell{timeout}{type-error}    \\
    \outcome{name-error}                           & \bvpcell{name-error}{success} & \bvpcell{name-error}{failure} & \bvpcell{name-error}{other} & \bvpcell{name-error}{name-error} & \bvpcell{name-error}{type-error} \\
    \outcome{type-error}                           & \bvpcell{type-error}{success} & \bvpcell{type-error}{failure} & \bvpcell{type-error}{other} & \bvpcell{type-error}{name-error} & \bvpcell{type-error}{type-error} \\
    \bottomrule
  \end{tabular}
  \caption{Number and percentage of all \bplsub programs generated in the experiments, for each combination of \plt execution outcome $p$ and Boogie verification outcome $b$ as in \autoref{tab:outcome-match}.
    A \colorbox{nocolbg}{red background} highlights the combinations
    that correspond to \emph{inconsistent} results that have been observed
    in the experiments.
  }
  \label{tab:checks-stats}
\end{table}

\subsection{Results: Consistency Checks}
\label{sec:results-consistency}

\autoref{tab:checks-stats} reports
the number and percentage of all 3 million programs
generated in the experiments
that resulted in each of \autoref{tab:outcome-match}'s
possible combinations of \plt execution outcome $p$ and
Boogie verification outcome $b$.

\subsubsection{Correctness}
In the vast majority of cases,
Boogie's outcome was consistent with
the result of executing according to \bplsub's operational semantics.
In particular, all programs with name errors
or type errors were caught by Boogie's name resolution
or type checking modules.

Boogie successfully verified \n:[0]{.80262176469344800408}[100]| (i.e., $(\n{Execution-vs-Verification/success-success/Count} + \n{Execution-vs-Verification/loop-success/Count}) / (\n{Execution-Outcomes/success/Count} + \n{Execution-Outcomes/loop/Count})$)
of all (partially) correct programs according
to \plt execution (that is, $p$ outcomes \outcome{success} and \outcome{loop});
it also confirmed as incorrect nearly
\n:[0]{.99998351832761968690}[100]|
(i.e., $\n{Execution-vs-Verification/failure-failure/Count} /  \n{Execution-Outcomes/failure/Count}$)
of all incorrect programs according
to \plt execution ($p$ outcome \outcome{failure}).
The cases of programs involving timeouts are inconclusive,
but they are also only a tiny fraction of the total.
Overall,
\bcc was useful to thoroughly validate Boogie,
which behaved correctly in the vast majority of cases.

\subsubsection{Soundness}
\bcc did not expose any soundness failure of Boogie,
that is cases of incorrect programs ($p = \outcome{failure}$)
that Boogie passes as correct ($b = \outcome{success}$).
While it is known that SMT solvers---including Z3---do
sometimes suffer from soundness bugs,
exposing them requires bespoke constraints~\cite{DBLP:journals/pacmpl/Winterer024},
which are unlikely to be generated by \bcc's
grammar during purely random enumeration.
Furthermore, Boogie's verification condition generation
algorithm adds a layer of indirection between
the Boogie program and the constraints sent to the SMT solver,
which may complicate triggering soundness bugs through Boogie
without involving features, such as triggers, that
directly interact with the underlying solver~\cite{BugariuTM21}.
Despite these limitations,
\bcc's testing remains useful to increase our confidence that
Boogie is generally sound.

\subsubsection{Completeness}
Our experiments found \n:{65347}
($\n{Execution-vs-Verification/success-failure/Count} + \n{Execution-vs-Verification/loop-failure/Count}$) cases
of \emph{completeness} failure;
that is,
Boogie rejected as (possibly) incorrect ($b = \outcome{failure}$)
about
\n:[0]{0.19733293874999622528}[100]|
(i.e.,
$\n:{65347} / (\n{Execution-Outcomes/success/Count} + \n{Execution-Outcomes/loop/Count})$)
of all correct programs
(\plt $p$ outcomes \outcome{success} or \outcome{loop}).
These results are largely consistent with how Boogie is designed:
since it aims at being a sound implementation of deductive program verification
(which is undecidable in general),
it will necessarily incur cases of incompleteness.
Still, they also further demonstrate \bcc's practical usefulness
in stress-testing a verifier to better reveal its capabilities and limitations.

\autoref{subsec:checking-consistency}
introduced the distinction between \emph{annotation} and \emph{reasoning}
incompleteness. How many of our experiments triggered each kind of incompleteness?
To answer this question, we tried to use
Boogie's support for loop invariant inference
as a proxy for distinguishing between reasoning and annotation incompleteness.
For each program $P$ where $b = \outcome{failure}$ but $p = \outcome{success}$ or \outcome{loop},
we ran Boogie on $P$ again with option \texttt{/infer:j}---which enables loop invariant inference.
If $P$ verifies successfully in this new run,
it suggests that it is a case of annotation incompleteness;
otherwise, we classify it as reasoning incompleteness.
We stress that this is an imperfect proxy,
which is, in general, neither sound nor precise:
\begin{enumerate*}
\item obviously, if $P$ still fails verification, it may just mean
  that Boogie's loop invariant inference could not find the ``right''
  loop invariant;
\item conversely, if $P$ passes verification, it may still be a case
  of reasoning incompleteness, where adding a loop invariant $J$
  unexpectedly helps establish that $A$ holds, even if $J$ is
  subsumed by $\neg C$ or $C$.
\end{enumerate*}

\begin{table}[!bt]
    \centering
    \setlength{\tabcolsep}{3pt}
  \begin{tabular}{l *{2}{rr}}
    \toprule
    \multicolumn{1}{c|}{} & \multicolumn{4}{c}{\textbf{Boogie outcome $b$}} \\
    \multicolumn{1}{c|}{} & \multicolumn{2}{c}{\emph{invariant inference}} & \multicolumn{2}{c}{} \\
    \multicolumn{1}{c|}{\textbf{\plt outcome $p$}} & \outcome{success} & \outcome{failure} & \outcome{success} & \outcome{failure} \\
    \midrule
    \outcome{success}                & \n{Execution-vs-Verification-Infer/success-success/Count} & \cellcolor{nocolbg}\n{Execution-vs-Verification-Infer/success-failure/Count} & \n{Execution-vs-Verification/success-success/Count} & \cellcolor{nocolbg}\n{Execution-vs-Verification/success-failure/Count} \\
    \outcome{loop}                   & \n{Execution-vs-Verification-Infer/loop-success/Count} & \cellcolor{nocolbg}\n{Execution-vs-Verification-Infer/loop-failure/Count} & \n{Execution-vs-Verification/loop-success/Count} & \cellcolor{nocolbg}\n{Execution-vs-Verification/loop-failure/Count} \\
    \bottomrule
  \end{tabular}
  \caption{How Boogie's inconsistency failures change if it uses loop invariant inference. The two rightmost columns are the same as in \autoref{tab:checks-stats}.
A \colorbox{nocolbg}{red background} highlights the combinations
that correspond to \emph{incompleteness} failures.}
  \label{fig:incompleteness-with-inv-inference}
\end{table}

With these caveats in mind,
let's have a look at
\autoref{fig:incompleteness-with-inv-inference},
which shows how the number of incompleteness inconsistencies
change if Boogie uses loop invariant inference.
The number of inconsistencies shrinks significantly:
only \n:[0]{.22734019924403568641}[100]|
(i.e.,
$(\n{Execution-vs-Verification-Infer/success-failure/Count} + \n{Execution-vs-Verification-Infer/loop-failure/Count}) / (\n{Execution-vs-Verification/success-failure/Count} + \n{Execution-vs-Verification/loop-failure/Count})$)
of spurious failures encountered in our experiments
remain even if loop invariant inference is enabled.
This
would seem to suggest that
completeness failures
often denote annotation incompleteness
rather than reasoning incompleteness.
On the other hand,
a nontrivial number of cases of actual annotation incompleteness
seem to remain
(those resistant to invariant inference).

In order to better understand
the effect of loop invariant inference,
we selected a small random sample
consisting of 40 Boogie programs
among those that 
\plt execution
confirmed as correct ($p = \outcome{success}$ or \outcome{loop}):
20 among those that
Boogie without loop invariant inference
flagged as incorrect ($b = \outcome{failure}$),
and Boogie with loop invariant inference
verified successfully ($b = \outcome{success}$);
and 20 among those that
Boogie flagged as incorrect ($b = \outcome{failure}$)
with or without loop invariant inference.
We manually inspected these 40 Boogie programs,
trying to determine whether they were cases of
annotation or reasoning incompleteness.
We found that 23 cases\footnote{
  Precisely, 14 cases that Boogie verifies only with \Bpl{/infer:j},
  and 9 cases that fail verification even with \Bpl{/infer:j}.
} out of 40
were actually
instances of \emph{reasoning} incompleteness,
since they all involved unreachable code guarded
by a condition that was identically false
(independent of any additional annotation).
Therefore, loop invariant inference
should not have any effect on whether
Boogie can verify these examples successfully.
In practice,
option \Rkt{/infer:j}
introduces several nontrivial changes
in how Boogie generates the verification conditions
of a program;
therefore, it is to be expected
that it can have an indirect, unpredictable
effect on the kinds of programs that can be automatically verified---it is a form of brittleness, also observed in related work~\cite{boogie-robustness-testing}.
In all,
we have reasons to believe that
a larger fraction of the completeness failures
encountered in \bcc's experiments
are actually indicative of \emph{reasoning} incompleteness---even though it is hard to get a precise estimate without
a very prohibitively time-consuming extensive manual analysis.
Regardless, our experiments
successfully demonstrated \bcc's practical applicability
to run large-scale thorough testing of the the Boogie intermediate verifier.

\begin{figure}[!bth]
  \begin{subfigure}[t]{0.5\textwidth}
\begin{lstlisting}[language=boogie]
procedure alwaysLoops() returns () {
  var G : bool;
  G := true;
  G := G;
  assert (0 < 1) && (1 == 1);
  while (G) {
    while (G) {
      assert !(true ==> false);
   }
   G := true;
  }
  //...
  assert false;
}
\end{lstlisting}
  \caption{The outer loop in this program never terminates.}
  \label{fig:examples:always-loops}
  \end{subfigure}
  \begin{subfigure}[t]{0.5\textwidth}
\begin{lstlisting}[language=boogie]
procedure neverLoops() returns () {
  var s: bool; var AE: bool;
  s := false; AE := false;
  while (s) {
    if ((-0 * 0 + -0) > 1) {
    } else {
      while (s) {
        AE := true;
        s := !(s && !false);
      }
    }
    s := true;
    if (0 >= -3) {
      if (false) {
      } else {
        assert (!(true && s) == AE);
      }
    }
    /* ... */ } }
\end{lstlisting}
  \caption{The outer loop in this program never executes.}
  \label{fig:examples:never-loops}
  \end{subfigure}
  \caption{Two programs generated by \bcc that expose incompleteness in Boogie.}
  \label{fig:examples:incompleteness}
\end{figure}

\subsubsection{Examples of Reasoning Incompleteness}
\autoref{fig:examples:incompleteness}
shows two programs generated by \bcc in our experiments\footnote{
  For readability, we simplified the full programs
  by removing parts that do not affect their behavior.
}
that showcase \emph{completeness} failures.
For space reasons we only shows these two examples,
which are, however, quite representative
of numerous other similar examples that exhibit the same behavior.

The outermost loop's body in
\autoref{fig:examples:always-loops}'s
program \Bpl{alwaysLoops}
(which has the same general structure as \autoref{fig:loop-with-inv}'s left program)
clearly never terminates,
since variable \Bpl{G} is initialized to \Bpl{true}
and never changes value.
While \plt execution
results in $p = \outcome{loop}$,
Boogie returns with $b = \outcome{failure}$,
flagging a violation of the \Bpl{assert false}
just after the outer loop.
According to our classification, this is a case of annotation incompleteness,
since Boogie correctly verifies the program
if it is given the additional information that \Bpl{G}
is always true (which can be retrieved by loop invariant inference).
Note that, in this case, the invariance of \Bpl{G} is path and flow insensitive:
in the loop, there is a single assignment to \Bpl{G} of the constant \Bpl{true}.

Conversely,
the outermost loop's body in
\autoref{fig:examples:never-loops}'s
program \Bpl{neverLoops}
clearly never executes,
since variable \Bpl{s} is initialized to \Bpl{false}
and immediately checked as the loop condition.
While \plt execution
results in $p = \outcome{success}$,
Boogie returns with $b = \outcome{failure}$,
flagging a violation of the assertion at the end of the loop body.
Since the outer loop condition is obviously identically \Bpl{false}
when it is first evaluated,
this is a case of reasoning incompleteness.
However, Boogie returns with $b = \outcome{success}$
if we enable the \Bpl{/infer:j} option,
even though a loop invariant
is not needed to determine that the loop never executes.
Even more unpredictably,
if we add more statements after the \Bpl{assert}
within the loop body,
then Boogie reports a (spurious) verification failure
regardless of whether loop invariant inference is
or isn't enabled.

\section{Conclusions}
This paper presented \bcc: a lightweight validation technique
to test deductive verifiers
such as the widely used Boogie.
\bcc is built atop a formal model
of a small, deterministic subset of the Boogie language,
consisting of a grammar, typing rules, and an executable operational semantics---encoded using the \plt semantic engineering framework.
In our experiments,
we used \bcc to thoroughly test Boogie for consistency;
while we found that Boogie's
output is correct and reliable in the vast majority of cases,
we did find a few interesting examples
that expose completeness failures.

\clearpage

\printendnotes[custom]

\end{document}